\documentclass[%
twocolumn,
groupedaddress,
amsmath,amssymb,
aps,
prb,
floatfix,superscriptaddress
]{revtex4-1}

\usepackage{mhchem}
\usepackage{amsmath}  
\usepackage{mathtools}
\usepackage{siunitx}
\usepackage{longtable}

\DeclareMathOperator*{\argmin}{argmin} 
\usepackage{epstopdf} 
\usepackage{graphicx}  
\usepackage{verbatim}  
\usepackage{color}   
\usepackage{subfigure} 
\usepackage{hyperref}  
\begin{document}

\title{Approximations in first-principles volumetric thermal expansion determination}

\author{Samare Rostami}
\email{samare.rostami@gmail.com}

\affiliation{European Theoretical Spectroscopy Facility, Institute of Condensed Matter and Nanosciences, Universit\'{e} catholique de Louvain, Chemin des \'{e}toiles 8, bte L07.03.01, B-1348 Louvain-la-Neuve, Belgium}

\author{Xavier Gonze}
\affiliation{European Theoretical Spectroscopy Facility, Institute of Condensed Matter and Nanosciences, Universit\'{e} catholique de Louvain, Chemin des \'{e}toiles 8, bte L07.03.01, B-1348 Louvain-la-Neuve, Belgium}

\begin{abstract}
In the realm of thermal expansion determination, the Quasiharmonic Approximation (QHA) stands as a widely embraced method for discerning minima of free energies across diverse temperatures such that the temperature dependence of lattice parameters as well as internal atomic positions can be determined. 
However, this methodology often imposes substantial computational demand, necessitating numerous costly calculations of full phonon spectra in a possibly many-dimensional geometry parameter space. 
Focusing on the volumetric thermal expansion only, the 
volume-constrained Zero Static Internal Stress Approximation (v-ZSISA) within QHA allows one to limit significantly the number of phonon spectra determinations to typically less than ten. The linear Gr\"uneisen approach goes even further with only two phonon spectra determinations to find the volumetric thermal expansion, but a deterioration of the accuracy of the computed thermal expansion is observed, except at low temperatures.
We streamline this process by introducing further intermediate approximations between the linear Gr\"uneisen and the v-ZSISA-QHA, corresponding to different orders of the Taylor expansion. The minimal number of phonon spectra calculations that is needed to maintain precise outcomes is investigated. The different approximations are tested on a representative set of 12 materials.
For the majority of materials, three full phonon spectra, corresponding to quadratic order, is enough to determine the thermal expansion in reasonable agreement with the v-ZSISA-QHA method up to 800 K. Near perfect agreement is obtained with five phonon spectra. This study paves the way to multi-dimensional generalizations, beyond the volumetric case, with the expectation of much bigger benefits.

\end{abstract}

\maketitle

\section{Introduction\label{sec:introduction}}

Thermal expansion of materials, impacting both scientific inquiry and practical applications, is intricately linked to the anharmonic behavior exhibited by solid materials. Understanding and predicting thermal expansion properties is essential for designing reliable devices that can operate effectively across a wide range of temperatures~\cite{Barron1980,wallace1972,grimvall1986,Barrera2005}.

The Quasiharmonic Approximation (QHA)~\cite{Dove1993,Lazzeri1998,Carrier2007,Allen2020} method is widely employed to investigate the temperature dependence of properties of weakly anharmonic solids. It takes into account crystal-structure-dependent phonon frequencies while ignoring dynamical phonon-phonon interactions. 
QHA operates under the assumption that phonon modes are harmonic, non-interacting, and solely influenced by the crystal lattice parameters and equilibrium internal atomic positions. 
The QHA free energy hypersurface, including harmonic phonon contributions, is a function of lattice parameters and internal atomic positions, for a given temperature, and must be minimized to yield the temperature-dependence of such degrees of freedom. 
In this approach, the free energy is the result of adding the QHA phonon free energy to the Born-Oppenheimer (BO) energy at zero temperature. For metals, a small electronic free energy correction might be added.  
The QHA phonon free energy is temperature dependent, and does not even vanish at zero temperature, due to zero-point motion.

Computing accurate phonon spectra from first principles is done routinely nowadays, thanks to Density Functional Perturbation Theory~\cite{Baroni1987, Gonze1997, Baroni2001, Gonze2005a} or frozen-phonon calculations ~\cite{Togo2015}. 
The robustness of the methodology is such that high-throughput phonon calculations are possible~\cite{Petretto2018a}.
This makes QHA particularly suitable for computing temperature-dependent thermodynamic and thermoelastic parameters under conditions where dynamic anharmonic effects are sufficiently small. 

Still, the computation of phonon spectra remains
relatively costly, and make the  QHA methodology computationally much more demanding than the straight determination of lattice parameters and internal atomic positions from the minimization of the BO energy. 
This inconvenience is reduced for materials where only one degree of freedom, the volume, govern lattice parameters (e.g. cubic lattices) while all internal degrees of freedom are fixed by symmetry. This is the case for most materials that have been studied using QHA in the past century. 
Indeed, the optimization of this unique degree of freedom involves only a one-dimensional search, with typically less than ten phonon spectrum calculations. 
Two- and even three-dimensional~\cite{Liu2018,Dangic2018} cases (needed to deal with e.g. tetragonal, rhombohedral, hexagonal and orthorhombic lattices) have been studied as well using the QHA. However, 
thermal expansion phenomena in crystals with incomplete symmetry-determined atomic arrangements usually depend on more than two or three degrees of freedom. 
They present computational challenges particularly in accurately modeling internal parameter (atomic position) changes with temperature.

Additional approximations are needed for such case with many degrees of freedom. 
Actually, despite its widespread use, the reliability of QHA has not been established for low-symmetry structures.

It is possible to decrease the effective number of degrees of freedom to be investigated in QHA thanks to the Zero Static Internal Stress Approximation (ZSISA), introduced by Allan and coworkers in 1996~\cite{Allan1996,Taylor1997,Taylor1999,Allen2020}. ZSISA streamlines the process by minimizing solely the BO energy with respect to atomic forces (also referred to as internal stresses in this work) for each external strain state, akin to QHA's consideration of harmonic phonon modes. 
This method enables the determination of external strains at each temperature with an error that is only second order in the thermal internal stresses. 
These thermal internal stresses are actually non-BO forces originating from the presence of phonons. 
Both thermal external stresses and thermal internal stresses will be referred to as thermal gradients in what follows.  
ZSISA may yield inaccurate internal strains, particularly at high temperatures or with significant zero-point energy, due to its neglect of such thermal internal stresses. 
Some studies, including those on wurtzite ZnO~\cite{Liu2018,Masuki2023} have demonstrated ZSISA's proficiency in predicting accurate external thermal expansions but have highlighted its shortcomings in internal atomic position determination, especially at elevated temperatures.
Actually, most of the QHA computations of anisotropic thermal expansion in the literature rely on ZSISA~\cite{Lichtenstein2000, Mounet2005,Carrier2007,Palumbo2017,Liu2018a,Ritz2018,Ritz2019,Li2021a,Brousseau2022}.

Further decrease in the computational cost is achieved when the focus is placed solely on volumetric expansion. In this case, the modified approach known as volume-constrained Zero Strain Internal Structure Approximation (v-ZSISA)~\cite{Skelton2015} is employed. 
In v-ZSISA, the cell volume (or hydrostatic strain) only is optimized on the basis of the free energy and determines volumetric expansion, while other parameters representing deviatoric strain and internal coordinates changes are optimized from the BO energy constrained as a function of volume. 
Not surprisingly, the prediction of thermal expansion anisotropy from v-ZSISA might be quite inaccurate. 

In v-ZSISA-QHA the phonon spectrum
is determined at many different volumes, typically between seven and twelve volumes~\cite{Nath2016}. 
Very frequently, for non-cubic crystals, the distinction between QHA and v-ZSISA-QHA is not made in the published results.  
Again, most of the QHA computations of volumetric thermal expansion in the literature rely on v-ZSISA
~\cite{Togo2010,Otero-de-la-Roza2011,Otero-de-la-Roza2011a,Li2011,Gupta2013,Togo2015,Skelton2015,Nath2016,Abraham2018}.

Recently, Masuki and coworkers~\cite{Masuki2023} have established (among other results) that if the set of parameters to be optimized is split in two, arbitrarily, with the free energy being optimized explicitly with respect to the first set (that might be the set of crystallographic parameters as in ZSISA, or might be restricted to the volume only as in v-ZSISA, but also might be more diverse than these two choices), while the values of the parameters in the second set (internal atomic coordinates in ZSISA or all parameters at fixed volume in v-ZSISA) are deduced from the BO energy, the first set determination agrees at the lowest order of the difference between the free  and BO energies with the result obtained from a full free energy optimization.

At variance with the reduction of the effective dimensionality of the problem thanks to ZSISA, v-ZSISA or their generalization, it is possible to reduce drastically the QHA computational load by 
considering a Taylor expansion of the Born-Oppenheimer energy to second order in the parameters to be optimized, and of the phonon free energy to first order in the parameters to be optimized. This is referred to as the linear Gr{\"u}neisen approach~\cite{Allen2020}. 
The derivative of the phonon free energy with respect to volume gives the well-known Grüneisen parameter~\cite{Gruneisen1912},
that can be computed from the mode-Gr{\"u}neisen parameters, i.e. the first derivative of each mode phonon frequency with respect to the volume.  
These concepts can be generalized to the derivative with respect to 
any geometrical parameter of the system, not simply the volume.
In the linear Grüneisen approach, the dimensionality of the problem is not reduced, but the scaling with respect to the dimensionality of the problem is much more manageable than in the QHA. 
The number of BO energy derivatives to be determined, typically elastic constants, linear internal stresses and interatomic force constants, scales quadratically with the dimension of the hyperspace in which the optimization is done, but can be determined with a practically linear scaling, 
thanks to the Hellman-Feynman theorem~\cite{Hellmann1937, Feynman1939}, 
while the number of phonon spectrum calculations is twice this dimension if Gr{\"u}neisen parameters are determined from symmetric finite differences.

The linear Gr{\"u}neisen method can also be specialized for the computation of the temperature-dependence of specific crystallographic parameters, instead of addressing all parameters.
Focusing on the volumetric thermal expansion, 
the linear Gr{\"u}neisen method only needs two phonon spectrum determinations. 
However, as mentioned earlier, its domain of validity is restricted.

The linear Gr{\"u}neisen approach nevertheless finds diverse applications, including the prediction of mode specific heat and the thermal expansion coefficient. 
Both the ZSISA-QHA and the linear Gr{\"u}neisen have been used in Ref.~\onlinecite{Brousseau2022} to predict zero-point lattice expansion (ZPLE) and its contribution to the total zero-point renormalization (ZPR) of the band gap energy in semiconductors and insulators. This latter study, focusing on zero-temperature effects, found
that using the ZSISA-QHA or the linear Gr{\"u}neisen
approach gives essentially identical results for a 
set of 22 materials belonging to the cubic or hexagonal crystallographic systems. 

It is known, and will be seen also in the present work, that the adequacy of the linear Gr{\"u}neisen approach is actually restricted to the low-temperature regime, lower than the Debye temperature, even for weakly anharmonic solids. 
Beyond that temperature, the thermal expansion coefficient from the linear Gr{\"u}neisen approach
quickly saturates and tends asymptotically to a constant, while the QHA thermal expansion coefficient usually continues to grow.

It would be desirable to join the concepts present in ZSISA-QHA and linear Gr{\"u}neisen methodologies, to enable the accurate computation of 
the temperature dependence in the case of a multi-dimensional parameter space. 
The linear Gr{\"u}neisen being too inaccurate, one
can wonder whether going to higher-order
expansions of the BO energy and phonon free energy
might restore the accuracy, still at a reasonable computational cost, provided the effective 
dimensionality of the space of the parameter is still reduced thanks to ZSISA.

A first step along such line has been done by Liu and Allen, in 2018~\cite{Liu2018a}. For ZnO, that has three degrees of freedom (two lattice parameters, and one internal atomic coordinate),
they examine the expansion of the BO energy to the third order, and the expansion of the phonon free energy to the second order, and gauge the accuracy
of the linear Gr{\"u}neisen method, that corresponds to an expansion of the BO energy to second order and the phonon free energy to first order. They also evaluate ZSISA.

We will also examine different orders of expansion, at this stage only focusing on the ability to 
obtain the volumetric thermal expansion with sufficient precision, while keeping in mind the many-dimensional perspective.
We will denote the linear Gr{\"u}neisen method as ``E2Vib1", with the first number being related to the BO energy expansion order, and the second one to the phonon free energy expansion order.
As the BO energy computation is much cheaper than phonon calculations, the obvious first step to improve upon the linear Gr{\"u}neisen is to drop any approximation for the BO energy, while keeping the minimal order for the phonon free energy, namely, first order. 
Such an approach is denoted E$\infty$Vib1, and requires two phonon spectrum calculations for each degree of freedom (lowest finite-difference approach), like the linear Gr{\"u}neisen method. 
Its cost is practically the same as the linear Gr{\"u}neisen method.
Using a quadratic Taylor expansion of the phonon free energy, still without any approximation for the BO energy, gives the E$\infty$Vib2 method. 
Such an expansion might be centered on the volume giving the BO energy minimum or around a larger volume, since volume expansion is expected. 
Including a fourth-order Taylor expansion of the phonon free energy gives the E$\infty$Vib4 method, also with different possibilities for choosing the volume around which the Taylor expansion is made.

In this work, the efficacy of the different approximations is tested on a set of twelve different materials representing various space groups, including MgO, Si, GaAs, Al, Cu, ZnO, GaN, AlN, \ce{YAlO3}, Bi, \ce{CaCO3}, and \ce{ZrO2}. 
This extensive analysis provides valuable insights into the effectiveness and limitations of these methods in capturing thermal expansion behavior. 
The temperature-dependent free energy and equilibrium volume are obtained, as well as the thermal expansion, from 0 K to 800 K. 
The E$\infty$Vib2 method is already doing an excellent job in the examined temperature range, while nearly perfect agreement is obtained for E$\infty$Vib4. Exceptions are discussed.

As mentioned earlier, the QHA is considered as a reference in the present work, that examines approximations to it, and establishes the trade-off between computational efficacy and accuracy.
So, dynamical phonon-phonon effects are neglected. 
Going beyond the QHA in order to include the latter effects has been the subject of several recent studies, that are now mentioned briefly, for sake of completeness. 
Deviations from harmonic behavior in atomic vibrations are clearly present, such as phonon broadening or frequency shifts at fixed volume. Molecular Dynamics (MD) simulations, the Self-Consistent Phonon (SCP) theory~\cite{Hooton1958,Masuki2022,Masuki2022a} and the Stochastic Self-Consistent Harmonic Approximation (SSCHA)~\cite{Errea2014a,Paulatto2015} are some of the approaches used to address such anharmonicity.
Including dynamical anharmonicity proved crucial to correctly describing the negative thermal expansion of ScF$_3$~\cite{Roekeghem2016,Oba2019}.
At variance, the one of ZrW$_2$O$_8$ is well described without dynamical anharmonicity~\cite{Vila2018}.

The structure of this paper is as follows. In Sec.\ref{sec:free_energy}, the free energy and its Taylor expansion are detailed, defining consecutively  the QHA, the ZSISA, the v-ZSISA,  the linear Gr{\"u}neisen method, the linearization of the vibrational contribution, and higher-order terms in the vibrational expansion. 
Thermal expansion is the focus of Sec.\ref{sec:thermal_expansion}, as it can be numerically computed more accurately using the entropy than by temperature-based finite-difference formulas.
The materials are presented in Sec.\ref{sec:materials}, and computational details in Sec.\ref{sec:computational_details}. Our results are presented in Sec.\ref{sec:results}, before the concluding section, Sec.\ref{sec:conclusion}.

\section{The free energy and its Taylor expansion\label{sec:free_energy}}

\subsection{The free energy and the corresponding optimized geometry}
Crystallographic parameters (lattice parameters, cell angles, and internal atomic positions) will be denoted by $C_\gamma$, where $\gamma$ runs from $1$ to 
$N_\textrm{C}$. 
Without considering symmetries, $N_\textrm{C}$ equals $6+3N_{\textrm{at}}-3$, where $6$ is the number of macroscopic crystallographic parameters and $N_{\textrm{at}}$ is the number of atoms in the primitive cell,
so that $3N_{\textrm{at}}-3$ parameters come from the internal degrees of freedom, excluding the global translations of the crystal. 
In practice, the number of independent crystallographic parameters is much lower, due to symmetries. 
One might also generalize this definition by incorporating magnetic variables into the set of crystallographic parameters, thanks
e.g. to constrained-DFT~\cite{Dederichs1984,Gonze2022e}.
The vector of all parameters $C_\gamma$ will be denoted as $\underline{C}$.
One is looking at the temperature dependence of these crystallographic parameters $C_\gamma(T)$ or $\underline{C}(T)$.

While lattice parameters and angles are well-defined macroscopic parameters, the internal  positions are 
averages of the values present in the macroscopic number of cells forming the solid.
In the present work, it will be supposed that 
the fluctuations of the atomic positions in each cell are around such unique average value. This hypothesis excludes the
treatment of a material
in which several local configurations with degenerate (or quasi-degenerate) energies
are present, and the system would jump from some local configuration to another one over time.
One supposes also that these crystallographic parameters can be modified smoothly by applying (in the computational setting) external stresses and internal forces, the latter being applied on a whole sublattice associated to some atomic position average value. 

The crystallographic parameters $C_\gamma(T)$, are optimized to minimize the free energy $F[C_\gamma,T]$:\begin{align}
F(T) &=\min_{\{C_\gamma\}} [F(C_\gamma,T)],\label{eq:1}\\
C_\gamma(T) &= \argmin_{\{C_\gamma\}} [F(C_\gamma,T)].\label{eq:2}
\end{align}

Equivalently, the temperature dependence of the parameters
$C_\gamma(T)$ is implicitely determined 
by the conditions for the minimization of  Eq.(\ref{eq:1})
\begin{align}
\left.\frac{\partial F}{\partial C_\gamma} \right|_{\underline{C}(T)} = 0. \label{eq:3}
\end{align}
 The free energy comprises several parts: the BO internal energy at 0 K (that is obviously temperature independent), the vibrational (phonon) part of the free energy, and corrections, due to e.g. electronic entropy and to coupled electron-phonon effects. 
 For insulators, they can be usually neglected. So, at this stage,
 
\begin{align}
 F(\underline{C},T)=
 E_{\textrm{BO}}(\underline{C})+F_{\textrm{vib}}(\underline{C},T).\label{eq:4}
\end{align}

For the purpose of the current work, it is supposed that $E_{\textrm{BO}}(C_\gamma)$ can be computed from first principles in a negligible amount of time compared to that needed for $F_{\textrm{vib}}(C_\gamma,T)$, which can also be computed from first principles. Additionally, it is assumed that the gradients of $F_{\textrm{vib}}(C_\gamma,T)$ with respect to the $C_\gamma$ parameters are not directly available, although they can be computed from finite differences.  Each $F_{\textrm{vib}}(C_\gamma,T)$ calculation, for a different set of 
$C_\gamma$  must be carefully planned. 

Plugging  Eq.(\ref{eq:4}) into Eq.(\ref{eq:3}) delivers a more explicit condition for the determination of 
$\underline{C}(T)$,
\begin{align}
	-\frac{\partial E_{\textrm{BO}}}{\partial C_{\gamma}}\Big|_{\underline{C}(T)}=
	\frac{\partial F_{\textrm{vib}}}{\partial C_{\gamma}}\Big|_{\underline{C}(T)}.
	\label{eq:5}
\end{align}
The right-hand side of Eq.(\ref{eq:5}) is termed the thermal gradient at position $\underline{C}(T)$.
If $\gamma$ is related to a cell parameter, $\partial F_{\textrm{vib}}/\partial C_{\gamma}$ is related to a stress.
If $\gamma$ is related to an atomic position, $\partial F_{\textrm{vib}}/\partial C_{\gamma}$ is related to a generalized (collective)force.

{
The electronic contribution to the free energy is also to be considered, in principle. Such electronic free energies 
($F_{\textrm{el}}$) coexist with vibrational free energies at various temperatures. Hence, Eq.(\ref{eq:4}) should be formulated as:
\begin{align}
	F(\underline{C},T)=E_{\textrm{BO}}(\underline{C})+F_{\textrm{vib}}(\underline{C},T)+F_{\textrm{el}}(\underline{C},T)\label{eq:4b}.
\end{align}
However, the electronic contribution
is much smaller than the phonon one, except for metals at very low temperature. 
In the present work , we have checked that it can be safely neglected {  for the materials under consideration and for the relevant temperature range}, not only for insulators/semiconductors, but for metals as well.}
{Nevertheless, in the case of $3d$, $4d$, or $5d$ metals with a high density of electronic states at the Fermi level, the electronic temperature could indeed have a notable impact, suggesting its consideration in the calculations.}

\subsection{The quasiharmonic approximation}
In the Quasi-Harmonic Approximation (QHA), atomic vibrations are assumed to be harmonic, but the vibrational frequencies are allowed to vary with crystallographic parameters and internal atomic positions. 
Accordingly, we make explicit this dependence by mentioning $\underline{C}$ as an argument of the frequencies,
with notation $\omega_{\textbf{q}\nu}(\underline{C})$, where
$\textbf{q}$ denotes the phonon wavevector
and $\nu$ the phonon branch.
Such computations are well-defined within a first-principle approach: 
interatomic force constants are 
provided as second-order derivatives of the
BO energy, irrespective of whether
the position of the atoms have
been relaxed under zero internal force, or if an internal force is present.

The vibrational free energy is then computed thanks to the usual Bose-Einstein statistics, that delivers the occupation number for each phonon mode $n_{\textbf{q}\nu}(\underline{C},T)$. The zero-point motion must also be included.
More precisely, the vibrational internal energy per unit cell is given by

\begin{align}
 U_{\textrm{vib}}(\underline{C},T)=
 \frac{1}{\Omega_{\textrm{BZ}}}
 \int_{\textrm{BZ}}
\sum_{\nu} \Big( \frac{1}{2}
+ n_{\textbf{q}\nu}(\underline{C},T) \Big)
\hbar\omega_{\textbf{q}\nu}(\underline{C})d\textbf{q},
 \label{eq:6_}
\end{align}
where the Brillouin zone volume 
$\Omega_{\textrm{BZ}}$ is related to the
primitive cell volume $\Omega_0$ by
$\Omega_{\textrm{BZ}}=\frac{(2\pi)^3}{\Omega_0}$.
The phonon frequencies 
$\omega_{\textbf{q}\nu}$ do not directly 
depend on the temperature.
The phonon occupation numbers $n_{\textbf{q}\nu}$
are given by the Bose-Einstein statistics,
\begin{align}
 n_{\textbf{q}\nu}(\underline{C},T)=
 \frac{1}{e^{\frac{\hbar \omega_{\textbf{q}\nu}(\underline{C})}{k_\textrm{B}T}}-1}.
 \label{eq:7_}
\end{align}
Similarly, the vibrational free energy per unit cell is written 
\begin{align}
 F_{\textrm{vib}}(\underline{C},T)&=
 \frac{1}{\Omega_{\textrm{BZ}}}
 \int_{\textrm{BZ}}
\sum_{\nu} 
\nonumber\\
&\Big( \frac{\hbar\omega_{\textbf{q}\nu}(\underline{C})}{2}
+ k_\textrm{B}T
\ln
\big(1-
e^{\frac{\hbar \omega_{\textbf{q}\nu}(\underline{C})}{k_\textrm{B}T}}\big)
\Big)
d\textbf{q},
 \label{eq:8_}
\end{align}
and the entropy per unit cell is
\begin{align}
 S_{\textrm{vib}}(\underline{C},T)&=
 -\frac{dF_{\textrm{vib}}}{dT}\Big|_{\underline{C}}=
 \frac{k_\textrm{B}}{\Omega_{\textrm{BZ}}}
 \int_{\textrm{BZ}}
\sum_{\nu} 
\nonumber\\
&\Big(- 
\ln
\big(1-
e^{\frac{\hbar \omega_{\textbf{q}\nu}(\underline{C})}{k_\textrm{B}T}}\big)
+ n_{\textbf{q}\nu}\frac{\hbar \omega_{\textbf{q}\nu}(\underline{C})}{k_\textrm{B}T}
\Big)
d\textbf{q}.
 \label{eq:9_}
\end{align}

One can check that 
\begin{align}
 F_{\textrm{vib}}(\underline{C},T)&=
 U_{\textrm{vib}}(\underline{C},T)
 -T S_{\textrm{vib}}(\underline{C},T).
 \label{eq:10_}
 \end{align}

The vibrational free energy Eq.(\ref{eq:8_}) contributes to the total free energy, following Eq.(\ref{eq:4}). Its
gradient enters Eq.(\ref{eq:5}).
The knowledge of entropy, Eq.(\ref{eq:9_}), will be shown later
to allow more accurate numerical thermal expansion determination.

\subsection{The v-ZSISA approximation}

According to the v-ZSISA approximation, the total free energy is minimized at fixed volume, while for such fixed volume, the BO energy
is minimized to find the values of the residual degrees of freedom (lattice parameters, angles, and internal reduced positions).

In this approach, one obtains the volumetric temperature dependence of free energies, $F(V,T)$. 
By minimizing the free energies with respect to the volume one obtains
a temperature-dependent volume.
Such minimization is typically done 
by fitting an equation of state (EOS) to the free energy data at each temperature.

{
Explicitly, the volume at temperature $T$, $V(T)$, is such that the pressure vanishes for that volume, the pressure being minus the derivative of the free energy with respect to volume,
\begin{align}
	P=-\frac{\partial F}{\partial V}\Big|_{V,T}.
	\label{eq:pressure}
\end{align}
So,
\begin{eqnarray}
	0&=&P(V(T))=
    -\frac{\partial F}
         {\partial V}\Big|_{V(T),T}
         \nonumber
 \\
&=& -\frac{\partial E_{\textrm{BO}}}{\partial V}\Big|_{V(T)}-
	\frac{\partial F_{\textrm{vib}}}
         {\partial V}\Big|_{V(T),T}
                  \nonumber
 \\
    &=&P_{\textrm{BO}}(V(T))+P_{\textrm{vib}}(V(T),T),
	\label{eq:zero_pressure}
\end{eqnarray}
where we have introduced the Born-Oppenheimer pressure $P_{\textrm{BO}}$ and the vibrational (or thermal) pressure $P_{\textrm{vib}}$, that must cancel each other at the equilibrium volume for the given temperature.
}

Given that the v-ZSISA-QHA requires phonon calculations at multiple volumes, it can be computationally demanding, even if much less so than the QHA or ZSISA-QHA. Therefore, we aim to employ various approximations that can achieve comparable accuracy while reducing the computational time required for these calculations.

\subsection{The linear Gr{\"u}neisen theory (E2Vib1)}
Expanding $E$ to second order and $F_{\text{vib}}$ to the first order around the optimized BO geometry at T = 0 K
(denoted $\underline{C}_{\textrm{BO}}$), is referred to by Liu and Allen\cite{Liu2018} as the linear Gr{\"u}neisen theory. 
As mentioned in the introduction,
the notation for the linear Gr{\"u}neisen theory in the current work is ``E2Vib1'', i.e. second order for the Taylor expansion of the BO energy
and first order for the Taylor expansion of the vibrational terms.
This method makes it possible, given current computational capabilities, to treat the temperature dependence of crystallographic parameters and internal degrees of freedom in a reasonably high-dimensional space.

The $\underline{C}_{\textrm{BO}}$ is determined by minimizing $E_{\textrm{BO}}$, 
\begin{align}
\frac{\partial E_{\textrm{BO}}}{\partial C_{\gamma}} \Big|_{\underline{C}_{\textrm{BO}}}=0.
\label{eq:13N}
\end{align}
Then, one expands the BO energy to second order in a Taylor series, with
$\Delta C_{\gamma}=	C_{\gamma}-C_{\textrm{BO}\gamma}$,
\begin{align}
E_{\textrm{BO}}(\underline{C})=&E_{\textrm{BO}}(\underline{C}_{\textrm{BO}})+\frac{1}{2}\sum_{\gamma \gamma'}
\Delta C_{\gamma}\Delta C_{\gamma'} 
\frac{\partial^2E_{\textrm{BO}}}{\partial C_{\gamma}\partial C_{\gamma'} }
\Big|_{\underline{C}_{\textrm{BO}}}.
\label{eq:11_}
\end{align}
This Taylor series has a constant term and a quadratic one, but no linear one, due to Eq.(\ref{eq:13N}). 
The vibrational free energy is stopped at first order:
\begin{align}
	F_{\textrm{vib}}(\underline{C},T) &= F_{\textrm{vib}}
 (\underline{C}_{\textrm{BO}},T)
	+&\sum_{\gamma} \Delta C_{\gamma} . 
\frac{\partial F_{\textrm{vib}}}
     {\partial C_{\gamma}}
\Big|_{\underline{C}_{\textrm{BO}},T}.
\label{eq:12_}
\end{align}
Plugging Eqs.(\ref{eq:11_}) and (\ref{eq:12_}) inside Eq.(\ref{eq:5}), at $C_{\gamma} = C_{\textrm{BO} \gamma}$, delivers:
\begin{align}
	\sum_{ \gamma'}
	\frac{\partial^2E_{\textrm{BO}}}{\partial C_{\gamma}\partial C_{\gamma'} } 
	\Big|_{\underline{C}_{\textrm{BO}}} \Delta C_{\gamma'} (T)
	=-\frac{\partial F_{\textrm{vib}}}{\partial C_{\gamma}} \Big|_{\underline{C}_{\textrm{BO}},T}.
\end{align}
The temperature-dependent $C_{\gamma} (T)$ is thus obtained from:
\begin{align}
	C_{\gamma} (T)&=
 C_{\textrm{BO}\gamma}& 
 \nonumber\\
 &+\sum_{ \gamma'}& \left(
	\frac{\partial^2E_{\textrm{BO}}}{\partial C_{\gamma}\partial C_{\gamma'} } 
	\Big|_{\underline{C}_{\textrm{BO}}} \right)^{-1}
	\left(-\frac{\partial F_{\textrm{vib}}}{\partial C_{\gamma}} \Big|_{\underline{C}_{\textrm{BO}},T}\right).
\end{align}
The second-order derivative of the Born-Oppenheimer energy is directly linked to elastic constants and phonon frequencies at $\Gamma$ for the optimized BO geometry, and can be obtained thanks to DFPT, 
even for a large set of 
geometrical parameter, in a much smaller CPU time than the full phonon band structure calculation. 
{
Indeed, the ratio between these two calculations is, roughly speaking, proportional to the number of wavevectors needed to sample the irreducible Brillouin Zone in order to obtain the full phonon band structure.
As an example, we can consider \ce{ZrO2}, a system comprising 12 atoms (the largest in our set) and of notable complexity. The vector space of atomic displacements is 36-dimensional, and the size of the dynamical matrices at the different wavevectors is $36\times 36$. Using DFPT, without symmetries, such dynamical matrix at $\Gamma$ is obtained by considering 36 perturbations. Linear combinations of these 36 perturbations span the whole vector space of atomic diplacements. Their linear response is a by-product of such DFPT calculations, whatever the collective displacement, by linear superposition. Owing to symmetries inherent to the system, this number is reduced to only 9 irreducible atomic displacements, to which the computation of the electric field perturbations (3) and strain perturbations (6) has to be added, hence 18 DFPT calculations. Let us consider now the computation of a full phonon band structure. A $4\times 4\times 4$ q-point grid is used. Also accounting for the symmetries of the system, the total number of perturbations to be considered in DFPT to generate the $36\times 36$ dynamical matrices for all q-points in the irreducible part of the Brillouin Zone increases to 324 (without explaining how the computer code determines this number). Consequently, the computational time required to compute all these perturbations is approximately 20 times greater than that needed for calculations solely at the gamma point.}

The
change of $F_{\textrm{vib}}$ with respect to the different parameters
is obtained by finite differences.  
The temperature enters in 
$\frac{\partial F_{\textrm{vib}}}{\partial C_{\gamma}} \Big|_{\underline{C}_{\textrm{BO}},T}$
only through the Bose-Einstein statistics, that determines the temperature-dependent phonon occupation numbers.
This makes the linear 
Gr{\"u}neisen theory applicable for 
the computation of the temperature-dependence of a large number of parameters.
The computational load is proportional to the time for one full phonon spectrum times twice the number of degrees of freedom $N_\textrm{C}$.

In cases where the dependence is solely on volume, the temperature-dependent volume in the E2Vib1 approach can be determined by:
\begin{align}
	V(T)=V_{\textrm{BO}}+\left(
	\frac{d^2E_{\textrm{BO}}}{d V^2 } 
	\Big|_{V_{\textrm{BO}}} \right)^{-1}
	\left(-\frac{dF_{\textrm{vib}}}{dV} \Big|_{V_{\textrm{BO}},T}\right),
	\label{eq:E2vib1}
\end{align}
where $V_{\textrm{BO}}$ represents the minimum energy volume, and 
$\frac{dF_{\textrm{vib}}}{dV} \Big|_{V_{\textrm{BO}},T}$ evaluated by a finite difference approach, possible with only two full phonon spectra calculations.
{
The bulk modulus being defined as
\begin{align}
	B={V}\frac{\partial^2 F}{\partial V^2}
    =-V\frac{\partial P}{\partial V},
	\label{eq:bulkmodulus}
\end{align}
one might equivalently express $V(T)$, Eq.(\ref{eq:E2vib1}), as
\begin{align}
	V(T)=V_{\textrm{BO}}
  \left(1+
 \left(
    B_{\textrm{BO}}
    \right)^{-1}
    P_{\textrm{vib}}(V_{\textrm{BO}},T)\right)
	,
	\label{eq:E2vib1_bulk}
\end{align}
where
\begin{align}
	B_{\textrm{BO}}={V_{\textrm{BO}}}\frac{\partial^2 E_{\textrm{BO}}}{\partial V^2}\Big|_{V_{\textrm{BO}}},
	\label{eq:bulkmodulus_BO}
\end{align}
}
an approximation to the QHA bulk modulus that is consistent with the E2Vib1 approach.

\subsection{Linearization of the vibrational contribution}
We now present a first intermediate level of approximation between the
QHA and the linear Gr{\"u}neisen approach.
Although we will test such intermediate
approximations only for the case of the volume dependence, the corresponding equations for the intermediate 
approximations will still be written
for the multidimensional case.
Moreover, we will consider an expansion around a general geometry, instead of the expansion around the BO geometry.

The idea of this first intermediate level, denoted as E$\infty$Vib1,
is to avoid the approximation of the
BO energy, but to keep the first-order
approximation of $F_{\textrm{vib}}$. 
Further approximations will retain progressively higher-order terms in the expansion of $F_{\textrm{vib}}$.

One expands the phonon free energy around a crystallographic configuration $\underline{C}^{\bullet}$ in a Taylor series with respect to $\Delta^{\bullet} C_{\gamma} = C_{\gamma} - C_{\gamma}^{\bullet}$, stopping at first order:

\begin{align}
	F_{\textrm{vib}}(\underline{C},T) = F_{\textrm{vib}}(\underline{C}^{\bullet},T)
	+\sum_{\gamma} \Delta ^{\bullet} C_{\gamma} . \frac{dF_{\textrm{vib}}}{dC_{\gamma}}\Big|_{\underline{C}^{\bullet},T}.
\label{eq:17_}
\end{align}
This method is denoted E$\infty$Vib1, first-order approximation for the vibrational contribution without approximation for the BO energy.
Inserting Eq.(\ref{eq:17_}) into Eq.(\ref{eq:5}), one gets:
 \begin{align}\label{eq:7}
 	-\frac{\partial E_{\textrm{BO}}}{\partial C_{\gamma}}\Big|_{\underline{C}(T)}=
 	\frac{\partial F_{\textrm{vib}}}{\partial C_{\gamma}}\Big|_{\underline{C}^{\bullet},T}	.
 \end{align}
 The right-hand side is not recomputed when determining $\underline{C}(T)$.
Equation (\ref{eq:7}) corresponds to the modified free energy minimization
 \begin{align}
 	F_{\textrm{E}\infty \textrm{Vib}1}(T) =&\min_{\underline{C}} 
 	\Big[~ E_{\textrm{BO}}(\underline{C})+  F_{\textrm{vib}}(\underline{C}^{\bullet},T)\nonumber\\
 +&	\sum_{\gamma}\Delta ^{\bullet} C_{\gamma} . \frac{\partial F_{\textrm{vib}}}{\partial C_{\gamma}}\Big|_{\underline{C}^{\bullet},T}\Big].
 	\label{eq:8} 
 \end{align}
The optimization with a fixed gradient is equivalent to an optimization under an external gradient, be it stress (or pressure) or forces.
 
\subsection{Higher-Order Terms in the vibrational Expansion}
\label{sec:IIF}
By retaining $E_{\textrm{BO}}$ without approximation and expanding the vibrational contribution into higher-order terms of the Taylor expansion, one can enhance the 
accuracy of the method.
Similar to E$\infty$Vib1, the phonon free energy is expanded around $\underline{C}^{\bullet}$, but this time, higher-order terms are considered.

\begin{align}
	F_{\textrm{vib}}[\underline{C},T] &= F_{\textrm{vib}}[\underline{C}^{\bullet},T]\nonumber\\ \nonumber
	+&\sum_{\gamma} \Delta ^{\bullet} C_{\gamma} . \frac{\partial F_{\textrm{vib}}}{\partial C_{\gamma}}\Big|_{\underline{C}^{\bullet},T}\\
	+&\frac{1}{2}\sum_{\gamma \gamma'} \Delta ^{\bullet} C_{\gamma} . \Delta ^{\bullet} C_{\gamma'} . \frac{ \partial ^2F_{\textrm{vib}}}{ \partial C_{\gamma} \partial C_{\gamma'}}\Big|_{\underline{C}^{\bullet},T}\nonumber\\
	+&\frac{1}{6}\sum_{\gamma \gamma' \gamma"} \Delta ^{\bullet} C_{\gamma} . \Delta ^{\bullet} C_{\gamma'}
	. \Delta ^{\bullet} C_{\gamma"} F^{'''}_{\textrm{vib},\gamma \gamma' \gamma"}\Big|_{\underline{C}^{\bullet},T}\nonumber\\
	+&\dots
\label{eq:20_}
\end{align}
We define E$\infty$Vib2 and E$\infty$Vib4 when terms up to the quadratic or fourth-order terms are included in the Taylor expansion Eq.(\ref{eq:20_}).

For the purpose of this paper, we aim to determine the dependency of the free energy on volume while keeping the other parameters optimized for each volume, following v-ZSISA. Consequently, the equations are rewritten in terms of $V$, $ V^{\bullet}$ and $\Delta^{\bullet} V = V-V^{\bullet}$ as follows:

\begin{align}
F_{\textrm{E}\infty \textrm{Vib}1}(V,T) &= E_{\textrm{BO}}(V)+ F_{\textrm{vib}}(V^{\bullet})\label{eq:Einfvib1}\nonumber\\
&+\Delta^{\bullet} V \frac{dF_{\textrm{vib}}}{dV}\Big|_{V^{\bullet},T},
\end{align}
\begin{align}
F_{\textrm{E}\infty \textrm{Vib}2}(V,T) &= E_{\textrm{BO}}(V)+ F_{\textrm{vib}}(V^{\bullet})\label{eq:Einfvib2}\nonumber\\
&+\Delta^{\bullet} V \frac{ \partial F_{\textrm{vib}}}{ \partial V}\Big|_{V^{\bullet},T}
\nonumber\\
&+\frac{1}{2} (\Delta^{\bullet} V)^2 \frac{ \partial ^2F_{\textrm{vib}}}{ \partial V^2}\Big|_{V^{\bullet},T},
\end{align}
\begin{align}
F_{\textrm{E}{\infty} \textrm{Vib}4}(V,T) &= E_{\textrm{BO}}(V)+ F_{\textrm{vib}}(V^{\bullet})\label{eq:Einfvib4}\nonumber\\
&+\Delta^{\bullet} V \frac{\partial F_{\textrm{vib}}}{\partial V}\Big|_{V^{\bullet},T}\nonumber\\
&+\frac{1}{2} (\Delta^{\bullet} V)^2 \frac{\partial ^2F_{\textrm{vib}}}{\partial V^2}\Big|_{V^{\bullet},T}\nonumber\\
&+\frac{1}{6} (\Delta^{\bullet} V)^3\frac{\partial ^3F_{\textrm{vib}}}{\partial V^3}\Big|_{V^{\bullet},T}\nonumber\\
 &+\frac{1}{24} (\Delta^{\bullet} V)^4 \frac{\partial ^4F_{\textrm{vib}}}{\partial V^4}\Big|_{V^{\bullet},T}.
\end{align}

In practice, one computes first the BO energy at a large number of volumes, which is relatively cheap, then the temperature-dependent vibrational free energy for the minimal
set of volumes allowing to deduce its Taylor expansion up to the desired order, and uses it to produce the total free energies for the same set of volumes as the BO energy. Subsequently, an Equation of State (EOS) or a high-order polynomial is fitted for each temperature and minimization of the EOS allows one to identify the optimal volume corresponding to each temperature.

The first derivative is obtained by considering at least two volumes, and the second derivative necessitates a minimum of three volumes. For the fourth derivative, at least five volumes are necessary. To provide a more detailed breakdown, in the case of E2Vib1 and E$\infty$Vib1, only two phonon calculations are necessary, at volumes $V_{\textrm{BO}}-\Delta V$ and $V_{\textrm{BO}}+\Delta V$, where $\Delta V$ represents the chosen volume spacing. For E$\infty$Vib2, the process entails calculations at three volumes ($V_{\textrm{BO}}-\Delta V$, $V_{\textrm{BO}}$, $V_{\textrm{BO}}+\Delta V$). Finally, for E$\infty$Vib4, the calculations are extended to five volumes ($V_{\textrm{BO}}-2\Delta V$, $V_{\textrm{BO}}-\Delta V$, $V_{\textrm{BO}}$, $V_{\textrm{BO}}+\Delta V$, $V_{\textrm{BO}}+2\Delta V$). In Vib4, all derivatives are calculated from five points.

As a further refinement, note that the finite difference evaluations might not be optimal if centered on the minimized BO geometry, since one expects thermal expansion. 
Thus one might also shift the different volumes towards higher ones, replacing 
$V_{\textrm{BO}}$ by some $V^{\bullet}$.
For example, for one might consider 
E$\infty$Vib2 done with phonon evaluations
at volumes ($V_{\textrm{BO}}$, $V_{\textrm{BO}}+\Delta V$
and $V_{\textrm{BO}}+2 \Delta V$), as will be done later as well.

At this stage, free energies (BO energy plus Taylor-expanded phonon free energies) are obtained at specific volumes, and the subsequent task is to determine these across all volumes and then minimize them to find the optimum volume at each temperature. To accomplish this, a function can be fitted to the data, with common choices being equations of state (EOS) such as the Vinet equation or polynomial functions. Notably, in our investigations, the fourth-degree polynomial has proven to be a more effective choice, particularly in complex cases. This preference arises from the fact that an EOS like the Vinet one requires a higher number of free energy points to achieve convergence compared to the polynomial approach. Accordingly, for both the QHA and our approximation models, we employed a fourth-degree polynomial to derive the results.

Let us emphasize a fundamental distinction between such methodology and the traditional Quasi-harmonic Approximation (QHA) in terms of the number of phonon calculations required to determine vibrational free energies. 
The QHA requires BO energy and phonon calculations across all different volumes.
The Taylor-based $F_{\textrm{vib}}$ methodologies rely on cheap calculations to obtain 
$E_{\textrm{BO}}(V)$, while 
subsequent phonon calculations are then conducted on the minimal number of volumes to determine the derivatives of vibrational free energies within the Taylor expansions.

\section{Thermal Expansion}
\label{sec:thermal_expansion}
Beyond obtaining the optimized $C_{\gamma}(T)$ values for different temperatures, one is also interested in obtaining their thermal expansion coefficients, related to the derivative of the $C_{\gamma}$ parameters with respect to the temperature, divided
by their value at a reference temperature (or, alternatively, at the
temperature where the derivative is evaluated).
Let us suppose that the set of $C_{\gamma}(T_1)$ has been obtained for some temperature $T_1$.
$\frac{dC_{\gamma}}{dT}\big|_{T_1}$ could be obtained by finite difference with neighboring temperatures. 
However we have observed that it is numerically more accurate to obtain such a
derivative thanks to another approach, from the first-order derivative of the entropy with respect to
$C_{\gamma}$
and the second derivatives of the free energy
with respect to the set of parameters 
$\underline{C}$. The latter is already available as a function of such $\underline{C}$, as described in the previous section.
The entropy can be obtained easily and accurately
from first principles, see Eq.(\ref{eq:9_}),
and then fitted as a function of $\underline{C}$,
similarly to the free energy.

\subsection{Multi-dimensional Case}

Because $F$ is minimal at $C_{\gamma}(T)$, from Eq.(\ref{eq:3}), one has, for all $\gamma$ and all $T$, 

\begin{align} 
\frac{\partial F}{\partial C_{\gamma}}\Big|_{C_{\gamma}(T),T} = 0.
 \label{eq:24_}
\end{align}
Using the chain rule for the total derivative of
this equation yields 
\begin{align}
0=&
	\sum_{\gamma' }\frac{\partial^2 F}{\partial C_{\gamma} \partial C_{\gamma'}} \Big|_{C_{\gamma}(T),T}
 \frac{dC_{\gamma'}}{dT}\Big|_{T}
+
 \frac{\partial^2 F}{\partial C_{\gamma} \partial T } \Big|_{C_{\gamma}(T),T}.
 \label{eq:25_}
\end{align}
Hence,
\begin{align}
	\frac{dC_{\gamma'}}{dT}\Big|_{T}=&-
	\sum_{\gamma }\left(\frac{\partial^2 F}{\partial C_{\gamma} \partial C_{\gamma'}} \Bigg|_{C_{\gamma}(T),T} \right)^{-1}
	\frac{\partial^2 F}{\partial C_{\gamma} \partial T } \Big|_{C_{\gamma}(T),T}\nonumber\\
	=&
	\sum_{\gamma }\left(\frac{\partial^2 F}{\partial C_{\gamma} \partial C_{\gamma'}} \Bigg|_{C_{\gamma}(T),T} \right)^{-1}
	\frac{\partial S}{\partial C_{\gamma} } \Big|_{C_{\gamma}(T),T},
 \label{eq:26_}
\end{align}
where  $S=-\frac{\partial F}{\partial T}$ is the entropy.

\subsection{Volumetric thermal expansion } 
When only the volumetric thermal expansion is considered, Eq.(\ref{eq:26_}) simplifies to:
\begin{align}
	\frac{dV}{dT}\Big|_{T}=&
	\Bigg( \frac{\partial^2 F}{\partial V^2}\Big|_{V(T),T} \Bigg) ^{-1}
	\frac{\partial S}{\partial V  } \Big|_{V(T),T}.
 \label{eq:thermal2}
\end{align}

In the QHA approach, $\frac{\partial^2 F}{\partial V^2}$ can be obtained by analytically deriving the Equation of State (EOS) or the polynomial fit mentioned at the end of Section \ref{sec:IIF}. 
Additionally, by fitting the entropy $S$ as a function of volume ($V$) using a third-order polynomial, one can determine $\frac{\partial S}{\partial V}$.

In the approximation methods discussed in the previous section, similarly to the vibrational free energy, $S$ should be expanded in a Taylor series around a crystallographic configuration $\underline{C}^{\bullet}$ (in this case, $V^{\bullet}$) with the same order of expansion as the free energy.

The expansion of $\frac{\partial S}{\partial V}$ at $V_1 = V(T_1)$ is written as:
\begin{align}
	\frac{\partial S_{\textrm{Vib}1}}{\partial V} \Big|_{V_1,T_1}=& \frac{\partial S}{\partial V} \Big|_{V^{\bullet},T_1}
	\nonumber	\\
			\frac{\partial S_{\textrm{Vib}2}}{\partial V} \Big|_{V_1,T_1}=& \frac{\partial S}{\partial V} \Big|_{V^{\bullet},T_1}+\Delta^{\bullet} V_1
		\frac{\partial^2 S}{\partial V^2} \Big|_{V^{\bullet},T_1} \label{ds1}\\
			\frac{\partial S_{\textrm{Vib}4}}{\partial V} \Big|_{V_1,T_1}=& \frac{\partial S}{\partial V} \Big|_{V^{\bullet},T_1}+
        \Delta^{\bullet} V_1
		\frac{\partial ^2S}{\partial V^2} \Big|_{V^{\bullet},T_1} \nonumber\\
		+&\frac{1}{2}(\Delta^{\bullet} V_1)^2\frac{\partial ^3S}{\partial V^3} \Big|_{V^{\bullet},T_1}\label{ds2}\nonumber\\
		+&\frac{1}{6}(\Delta^{\bullet} V_1)^3\frac{\partial ^4S}{\partial V^4} \Big|_{V^{\bullet},T_1} 
\end{align}
 
Therefore, for the linear Grüneisen theory (E2Vib1), from Eq.(\ref{eq:E2vib1}), one has:
 
    \begin{align}
 \left.\frac{dV_{\textrm{E}2\textrm{Vib}1}}{dT}\right|_{T_1} &= 
\left(\left.\frac{\partial^2 E_{\textrm{BO}}}{\partial V^2}\right|_{V_{\textrm{BO}}}\right)^{-1}
\left(\left.\frac{\partial S}{\partial V}\right|_{V_{\textrm{BO}},T_1}\right)
 \end{align}
By substituting Eq.(\ref{eq:Einfvib1}), Eq.(\ref{eq:Einfvib2}), Eq.(\ref{eq:Einfvib4}), and Eq.(\ref{ds1}, \ref{ds2}) into Eq.(\ref{eq:thermal2}), the derivative of $V$ with respect to $T$ is defined as follows for the different other cases:

    \begin{align}
\left. \frac{dV_{\textrm{E}\infty\textrm{Vib}1}}{dT}\right|_{T_1} &= 
 \left(\left.\frac{\partial^2 E_{\textrm{BO}}}{\partial V^2}\right|_{V_1,T_1}\right)^{-1}
\left.\frac{\partial S_{\textrm{Vib1}}}{\partial V}\right|_{V_1,T_1}\label{eq:thermal_Einfvib1}
 \end{align}

    \begin{align}
 \left.\frac{\partial V_{\textrm{E}\infty\textrm{Vib}2}}{\partial T}\right|_{T1} =&
 \left( \left.\frac{\partial^2 E_{\textrm{BO}}}{\partial V^2}\right|_{V_1,T_1} 
 + \left.\frac{\partial^2 F_{\textrm{vib}}}{\partial V^2}\right|_{V^{\bullet},T_1}
\right)^{-1}
\nonumber\\
&\frac{\partial S_{\textrm{Vib2}}}{\partial V}
\Bigg|_{V_1,T_1}
\label{eq:thermal_Einfvib2}
\end{align}

    \begin{align}
\left. \frac{\partial V_{\textrm{E}\infty\textrm{Vib}4}}{\partial T}\right|_{T1} =& \left(
\left. \frac{\partial^2 E_{\textrm{BO}}}{\partial V^2}\right|_{V_1,T_1} + \left.\frac{\partial^2 F_{\textrm{vib}}}{\partial V^2}\right|_{V^{\bullet},T_1}\right.
\nonumber\\
             +&\left.\Delta^{\bullet} V_1\frac{\partial^3 F_{\textrm{vib}}}{\partial V^3}\right|_{V^{\bullet},T_1}
\nonumber\\
             +&\left.\left.\frac{1}{2}(\Delta^{\bullet} V_1)^2 \frac{\partial^4 F_{\textrm{vib}}}{\partial V^4}\right|_{V^{\bullet},T_1}\right)^{-1}
\left.\frac{\partial S_{\textrm{Vib4}}}{\partial V}\right|_{V_1,T_1}
\nonumber\\
\label{eq:thermal_Einfvib4}
 \end{align}

\section{Materials}
\label{sec:materials}

The present work focuses on the temperature dependence of the volume. The v-ZSISA-QHA is invoked to relax the internal coordinates and specific lattice parameters and crystallographic angles, determined from zero-temperature, volume-constrained, DFT calculations. 
However, in preparation for further studies that will include anisotropic thermal expansion and dependence of atomic positions on temperature,
materials with many more degrees of freedom
are also included in the test set, e.g. up to ZrO$_2$ with 13 degrees of freedom (4 for the lattice and 9 for the atomic positions). 
The test set covers five crystallographic systems, namely cubic, hexagonal, orthorhombic, rhombohedral, and monoclinic. Table~\ref{tab:cases}  presents our set of twelve materials, their crystallographic system, and the independent lattice parameters and angles.

\begin{table}
\caption{The twelve materials studied in the present work, listed according to their crystallographic system. The rightmost column specifies the possible relations between crystalline parameters and/or the values of their angles.}

	\begin{ruledtabular}
		\begin{tabular}{llcccc}
			\begin{tabular}{l}MgO, Si, GaAs\\ Cu, Al \end{tabular}& Cubic &\begin{tabular}{c}~$a=b=c$, \\$\alpha=\beta=\gamma=90^{\circ}$ \end{tabular}\\
   \hline
			ZnO, AlN, GaN &  Hexagonal & \begin{tabular}{c}$a=b\neq c $,\\ $\alpha=\beta=90^{\circ}$\\ $\gamma=120^{\circ}$ \end{tabular}\\
    \hline
			YAlO$_{3}$& Orthorhombic&  \begin{tabular}{c}$a \neq b \neq c $,\\$ \alpha=\beta=\gamma=90^{\circ}$ \end{tabular}\\
    \hline
			Bi, CaCO$_{3}$& Rhombohedral&  \begin{tabular}{c}$a=b=c$,\\$ \alpha=\beta=\gamma\neq 90^{\circ}$ \end{tabular}\\
    \hline
			ZrO$_{2}$& Monoclinic   &\begin{tabular}{c}$a \neq b \neq c $,\\$ \alpha=\gamma=90^{\circ} \neq \beta$ \end{tabular}
			\label{tab:cases}
		\end{tabular}
	\end{ruledtabular}
\end{table}

Further, in Table~\ref{tab:tab2}, we report, for each material, the specific space group, the theoretical DFT lattice parameter(s), the experimental lattice parameter(s), the number of degrees of freedom, the number of atoms in the primitive cell, and the number of internal degrees of freedom, as well as some planewave computational parameters, namely, the cut-off  kinetic energy and electronic and vibrational wavevector samplings in the Brillouin Zone.

MgO, Si, GaAs, Al and Cu, have a face-centered cubic (FCC) Bravais lattice, with rock-salt, diamond, zinc-blende and the FCC structures (the latter both for Cu and Al). For these five solids, there are no degrees of freedom associated with atomic positions.
\ce{ZnO}, \ce{AlN} and \ce{GaN} belongs to the hexagonal crystallographic system, space group $P6_3mc$, with the four-atom wurtzite structure. The inherent symmetries in this configuration lead to a single internal degree of freedom, specifically the distance between different elements in the z-direction. 
\ce{YAlO3} belongs to the orthorhombic crystallographic system and has 20 atoms in the primitive cell, with 7 internal degrees of freedom. The rhombohedral structures of Bi and \ce{CaCO3}, with 2 and 10 atoms in the primitive cells, respectively, have one internal degree of freedom.
The most complicated case, monoclinic \ce{ZrO2}, features a crystal structure with 12 atoms arranged in positions that result in 9 internal degrees of freedom. It possesses a lower symmetry, and belongs to the $P2_1/c$ space group.

\section{Computational details}
\label{sec:computational_details}

\subsection{First-principles calculations}

Ground-state BO energies have been obtained from Density Functional Theory (DFT) and phonon frequencies from Density-Functional Perturbation Theory (DFPT). Spin-orbit coupling was omitted from the calculations for all materials except for Bismuth (Bi). Norm-conserving pseudopotentials\cite{Hamann2013} sourced from the \verb|Pseudo-Dojo|\cite{VanSetten2018} web site were utilized, employing the GGA-PBEsol functional~\cite{Perdew2008}. A comparative study of the effect of the choice of functional, including the LDA and different GGA, has been presented in 
Ref.~\onlinecite{He2014}, showing that GGA-PBEsol 
achieves comparatively good accuracy with respect to experiment for both the determination of the lattice parameter and the determination of phonon frequencies. 

The optimization process for lattice parameters and relevant internal atomic coordinates ran until atomic forces reached values below \(10^{-5}\) Hartree/Bohr\(^3\) and stresses values
below \(10^{-7}\) Hartree/Bohr\(^3\). 
Furthermore, to minimize errors in the second-order derivative of the free energy, we ensure that the largest square of the residual wave function is maintained below $10^{-20}$ Ha$^2$ in the ground state calculations~\cite{Gonze2024}.

In order to obtain smooth total energies as a function of the volume, an energy cutoff smearing~\cite{Laflamme2016} of 1.0 Ha was consistently applied. 
The electronic wavevector sampling for the Brillouin zone and the energy cutoff for each material were selected to achieve residual error levels smaller than 1 meV/atom, and are mentioned in Table \ref{tab:tab2}. 
For metallic materials (Cu, Al) and also Bi, a broadening of \(\eta = 0.02\) Ha was introduced using the resmearing technique employing the Methfessel-Paxton (MP) method ~\cite{Miehlich1989,Verstraete2002,Gonze2024} to ensure convergence in wavevector sampling.

The effect of electronic temperature was disregarded due to its  negligible impact on the thermal expansion derived from vibrational free energy, having no discernible effect on the final computational outcomes (much less than 1\% relative change).

The ABINIT software package (v9.8.3.) \cite{Gonze2002,Gonze2020,Romero2020} was employed for the computations.
The phonon density of states (PHDOS) was determined utilizing the Gaussian method, with a DOS smearing value set to  \(4.5 \times 10^{-6}\)  Hartree. Furthermore, a frequency grid step of \( 1.0 \times 10^{-6}\)  Hartree was employed for PHDOS calculations. These adjustments in numerical accuracy were imperative for versions of ABINIT preceding v9.10. Notably, in ABINIT versions v9.10 and after, these parameter values are preset as defaults for calculations.

\begin{table*}
	\caption{For each material: space group (Hermann–Mauguin notation), lattice parameters obtained by DFT calculations or experimentally (room temperature), angles between primitive cell vectors (``$\angle$") determined by DFT and experiments, number of lattice degrees of freedom (lDOF), number of atoms within the primitive cell, number of internal degrees of freedom (iDOF), alongside computational parameters such as planewave energy cutoff ($E_{\textrm{cut}}$), electronic wavevector sampling (k-grid), and vibrational wavevector sampling (q-grid). }
\begin{ruledtabular}
	\begin{tabular}{l|cccccccccc}
		
		Material &  Group &DFT lattice (\AA) &Exp. lattice (\AA) & $\angle$($^\circ$) & lDOF &No. atoms & iDOF & $E_\textrm{cut}$(Ha) & k grid & q grid \\
		\hline
		\ce{MgO  }&$Fm\bar{3}m $&\begin{tabular}{c} a= 2.980\end{tabular} & a=2.978 \cite{Li2006}                       &$\alpha$= 60                  &1 &2&0&60& 8 $\times$8 $\times$8 & 8 $\times$8 $\times$8  \\
  
		\ce{Si   }&$Fd\bar{3}m $&\begin{tabular}{c} a= 3.840\end{tabular}  & a=3.840 \cite{2002}                 &$\alpha$= 60             &1 &2&0&50& 8 $\times$8 $\times$8 & 8 $\times$8 $\times$8  \\
  
	\ce{GaAs }&$F\bar{4}3m $&\begin{tabular}{c}a= 4.003\end{tabular}  & a=3.995\cite{Madelung1999}  &$\alpha$= 60 
 &1   &2&0&60& 8 $\times$8 $\times$8 & 8 $\times$8 $\times$8  \\	
 
	\ce{Al   }&$Fm\bar{3}m $&\begin{tabular}{c} a= 2.837\end{tabular}  &\begin{tabular}{c} a= 2.856~\cite{Wilson1941}\end{tabular}                       &\begin{tabular}{c}$\alpha$= 60 \end{tabular}                  &1 &1&0&45& 12$\times$12$\times$12& 12$\times$12$\times$12 \\
 
	\ce{Cu   }&$Fm\bar{3}m $&\begin{tabular}{c} a= 2.518\end{tabular} &\begin{tabular}{c} a = 2.556\cite{Ashcroft1976}\end{tabular}       &\begin{tabular}{c}$\alpha$= 60 \end{tabular}                 &1  &1&0&55& 12$\times$12$\times$12& 12$\times$12$\times$12 \\
 
\ce{ZnO  }&$P6_3mc$&\begin{tabular}{c} a= 3.227\\ c= 5.207 \end{tabular}  &\begin{tabular}{l} a= 3.250~\cite{Karzel1996} \\ c= 5.204\end{tabular}            &\begin{tabular}{c}$\alpha$= 90 \\  $\gamma$= 120\end{tabular}           &2&4&1&52& 7 $\times$7 $\times$5 & 7 $\times$7 $\times$5  \\

\ce{AlN  }&$P6_3mc$&\begin{tabular}{l} a= 3.113 \\ c= 4.983 \end{tabular}   &\begin{tabular}{l} a= 3.110~\cite{SCHULZ1977}\\c=4.980\end{tabular}     &\begin{tabular}{c}$\alpha$= 90 \\  $\gamma$= 120 \end{tabular} &2 &4&1&52& 8 $\times$8 $\times$5 & 8 $\times$8 $\times$5  \\ 

\ce{GaN  }&$P6_3mc$&\begin{tabular}{c} a= 3.184\\ c= 5.187 \end{tabular}  &\begin{tabular}{l} a= 3.190~\cite{SCHULZ1977}\\c=5.189\end{tabular}  &\begin{tabular}{c}$\alpha$= 90 \\  $\gamma$= 120 \end{tabular} &2 &4&1&50& 8 $\times$8 $\times$5 & 8 $\times$8 $\times$5  \\

\ce{YAlO3}&$Pnma  $&\begin{tabular}{c} a= 5.322\\ b= 7.358 \\ c= 5.158 \end{tabular}&\begin{tabular}{l} a= 5.330\cite{DIEHL1975}\\ b= 7.375 \\ c= 5.180 \end{tabular}&$\alpha$= 90              &3  &20&7&60& 4 $\times$3 $\times$4 & 4 $\times$3 $\times$4  \\

\ce{Bi   }&$R\bar{3}m  $&\begin{tabular}{c} a= 4.744\end{tabular}  &\begin{tabular}{l} a=4.726~\cite{Wyckoff1963}
\end{tabular}                        &\begin{tabular}{c}$\alpha^{\text{DFT}}$= 57.50\\$\alpha^{\text{Exp.}}$~\cite{Wyckoff1963}=57.32 \end{tabular}              &2 &2&1&40& 8 $\times$8 $\times$8 & 8 $\times$8 $\times$8  \\

\ce{CaCO3}&$R\bar{3}c  $&\begin{tabular}{c} a= 6.313\end{tabular}  &\begin{tabular}{c} a=6.344~\cite{wang2018}\end{tabular}                        &\begin{tabular}{c}$\alpha^{\text{DFT}}$= 46.58 \\$\alpha^{\text{Exp.}}$~\cite{wang2018}= 46.31\end{tabular}            &2  &10&1&56& 5 $\times$5 $\times$5 & 5 $\times$5 $\times$5  \\

\ce{ZrO2 }&$P2_1/c$&\begin{tabular}{c} a= 5.126\\ b= 5.209 \\ c= 5.293 \end{tabular}&\begin{tabular}{l}  a= 5.169\cite{McCullough1959}\\ b= 5.232 \\ c= 5.341 \end{tabular}&\begin{tabular}{c}$\alpha$= 90 \\ $\beta^{\text{DFT}}$= 99.58\end{tabular} &4&12&9&56& 4 $\times$4 $\times$4 & 4 $\times$4 $\times$4   
		\label{tab:tab2}                                                                                                                                     
	\end{tabular}
\end{ruledtabular}
\end{table*}

\subsection{Sets of volumes}

\begin{figure}[t!]	
	\includegraphics[width=0.48\textwidth]{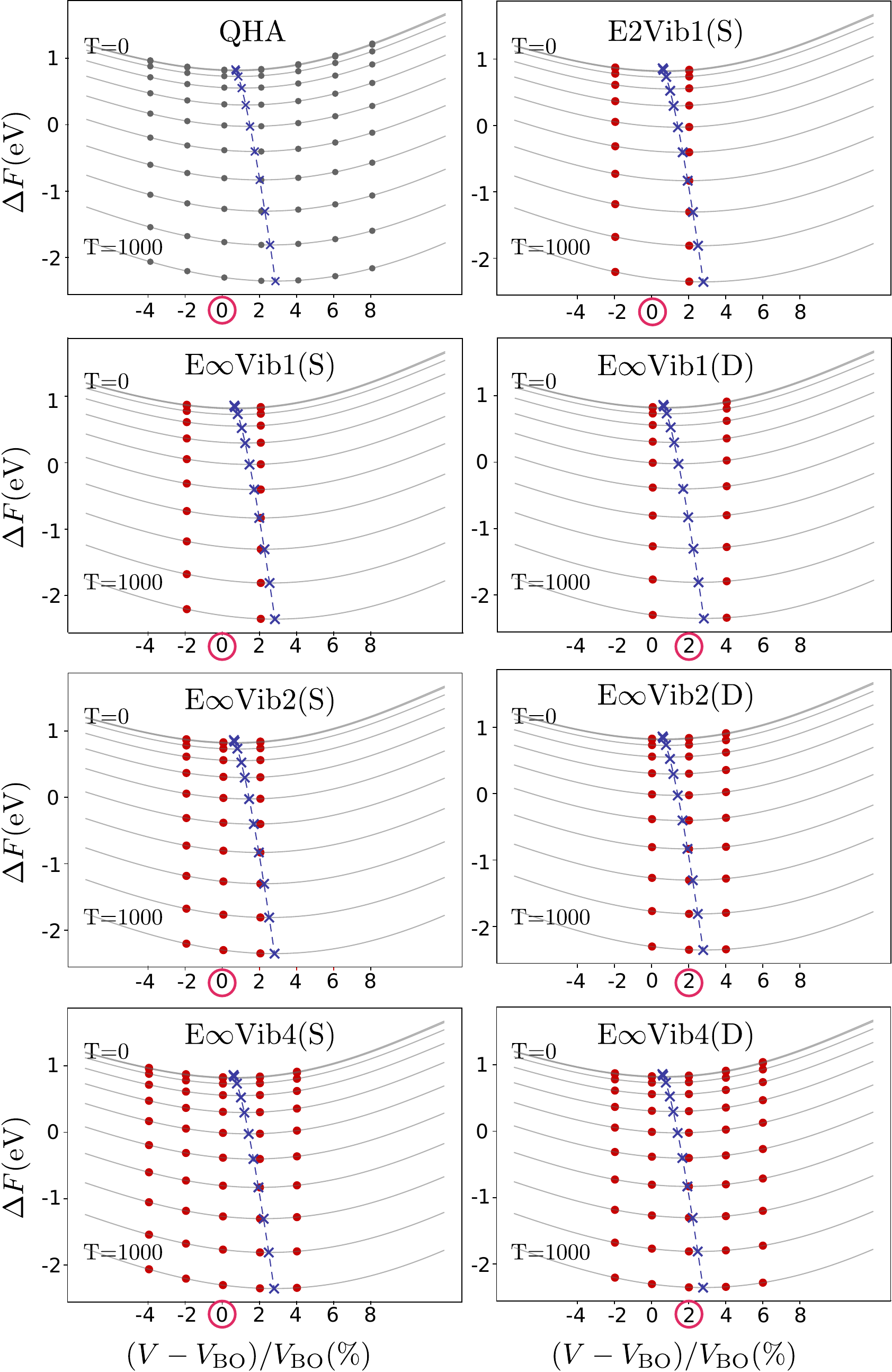}
		\caption{
  Illustration of the different approximations examined in the present study. 
  The free energy of ZrO$_2$ is obtained as a function of the percentage of volume changes at various temperatures. 
  In symmetric models (S), the reference volume in Eqs.(\ref{eq:Einfvib1}-\ref{eq:Einfvib4})  is fixed at the Born-Oppenheimer DFT optimal volume $V_{\textrm{BO}}$.  
  For displaced methods (D), the reference volume is shifted by 2\%. 
  On the horizontal axis, such shift is indicated by a red circles. Points represent values obtained through BO energy and phonon calculations. 
  The lines are interpolations from the approximation equations. 
  The top interpolating curve is the $T = 0$~K curve, whose free energy is lowered with respect to the BO one (not represented) by the zero-point motion energy. 
  The other curve goes to $T = 1000$~K by step of 100 K. 
  Crosses indicate the minima of each curve, for 11 temperatures
  going from 0~K to 1000~K by step of 100~K. 
  They are joined by the dashed line, providing the optimized values at much more than 11 temperatures. Indeed obtaining an 
  interpolating curve for a given temperature is easy to produce, as the compute-intensive calculations is the one of the phonon spectrum at a given volume. 
			}
		\label{fig:methods}
\end{figure}

In all instances, we maintained a consistent volume spacing of 2\% between different volumes, and included in the set of volumes used for each approximation the BO optimized minimum volume, $V_{\textrm{BO}}$, obtained in the Density Functional Theory (DFT) minimization process. 
The choice of 2\% was found effective in reducing the impact of numerical errors on phonon calculations.

At each volume, geometry optimization of the BO energy was performed, adapting all other degrees of freedom,
such as lattice parameters, angles, and ionic positions, while maintaining a fixed volume,
as required in the v-ZSISA-QHA approach.

To generate our reference v-ZSISA-QHA data, phonon calculations (and BO energy calculations) were conducted for a set of seven different volumes ranging from -4\% to +8\% of $V_{\textrm{BO}}$. Such a relatively low number of phonon spectra calculations for the QHA could be decided because of the accurate determination of phonons from ABINIT, without numerical fluctuations, as described in the previous subsection. However, the -4\% $V_{\textrm{BO}}$ volume was excluded for \ce{CaCO3} due to structural instability, with imaginary phonon modes. Additionally, the +10\% $V_{\textrm{BO}}$ volume was included for aluminum, copper, and bismuth to account for their considerable thermal expansion.
Temperature-dependent volumes and corresponding thermal expansion coefficients were determined from $T$=0 K to 1000 K.

In the approximation methods, BO energy calculations were made for the same volumes as those for QHA, while Density Functional Perturbation Theory (DFPT) calculations were executed at specific volumes corresponding to each approximation. Figure~\ref{fig:methods}, featuring \ce{ZrO2}, demonstrates the number of phonon calculations for each approximation, providing a visual representation of the associated computational demand.

Two distinct alternatives for selecting volume changes were considered: symmetric (S) and displaced (D). In symmetric choices, all volume changes were symmetrically selected around $V_{\textrm{BO}}$. 
For E2Vib1(S) and E$_{\infty}$Vib1(S), phonon spectra were obtained at $\pm2\%V_{\textrm{BO}}$.
 Similarly, in the case of  E$_{\infty}$Vib2(S)
 , volume changes included $\pm2\%V_{\textrm{BO}}$ and $V_{\textrm{BO}}$, 
 while for E$\infty$Vib4(S), volume changes encompassed  $\pm4\%V_{\textrm{BO}}$,  $\pm2\%V_{\textrm{BO}}$, 
 and $V_{\textrm{BO}}$.

Considering that usually the thermal expansion is positive, so that the optimum volume at each temperature exceeds $V_{\textrm{BO}}$, the reference volume in Eqs. (\ref{eq:Einfvib1})-(\ref{eq:Einfvib4}) was adjusted by setting $V^{\bullet}$ to the first volume change (+2\% $V_{\textrm{BO}}$) for
the displaced approximation, while in the symmetric approximation  
$V^{\bullet}$ was set to $V_{\textrm{BO}}$. 

 Consequently, volume changes were symmetrically selected around this new displaced volume, +2\% $V_{\textrm{BO}}$. This resulted in the selection of $V_{\textrm{BO}}$ and +4\% $V_{\textrm{BO}}$ for E$\infty$Vib1(D). 
 Similarly, $V_{\textrm{BO}}$, +2\% $V_{\textrm{BO}}$, 
 and +4\% $V_{\textrm{BO}}$ for E$\infty$Vib2(D), and -2\% $V_{\textrm{BO}}$, $V_{\textrm{BO}}$, +2\% $V_{\textrm{BO}}$, +4\% $V_{\textrm{BO}}$, and +6\% $V_{\textrm{BO}}$ for E$_{\infty}$Vib4(D). 
 Notably, the displacement approach was not applicable for E2Vib1 as the DFT energy must be minimized at $V_{\textrm{BO}}$.
The findings, see later, suggest that the displaced method is more effective for 
E$\infty$Vib2 and E$\infty$Vib1. 
With the exception of \ce{ZrO2}, the symmetric and displaced approximations in E$\infty$Vib4 are either identical or compatible.

\section{Results}
\label{sec:results}
 
\begin{figure}[!]	
		\includegraphics[width=0.45\textwidth]{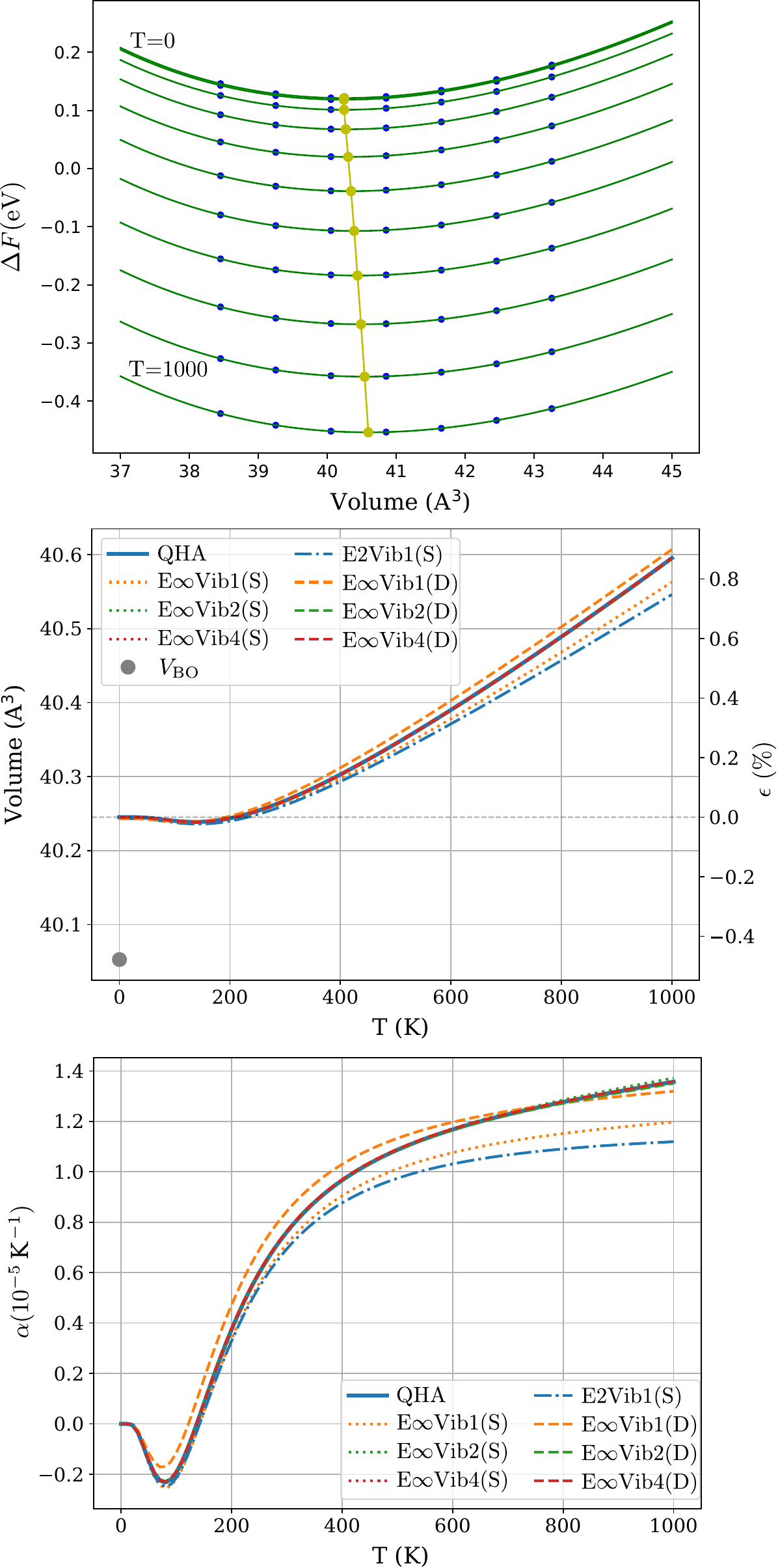}
	\caption{
 Detailed representation of the free energy, volume, and 
 thermal expansion coefficient of Si, for a temperature range from 0 K to 1000 K. In the top panel, free energies in QHA are charted. The minima of free energies are denoted by yellow points and lines. 
 The middle panel illustrates volume-temperature ($V$ vs. $T$) relationships for various approximations. Dashed lines represent results from the displaced selection of $V$, while dotted lines 
 depict results from symmetric selections. Solid lines denote QHA, and dot-dashed lines represent the E2Vib1 method. Right ticks indicate the percentage of volume change relative to the QHA volume at $T = 0$~K,
 $\epsilon=(V-V_{\textrm{QHA}}(T=0
 \textrm{~K}))/V_{\textrm{QHA}}(T=0 \textrm{~K})$. The gray dot highlights the optimum value from BO DFT minimization, $V_{\textrm{BO}}$. In the bottom panel, thermal expansion coefficients are presented using different approximations, with the same line definitions as in the middle panel.}
\label{fig:Si}
\end{figure}

\begin{figure}[!]	
	\includegraphics[width=0.45\textwidth]{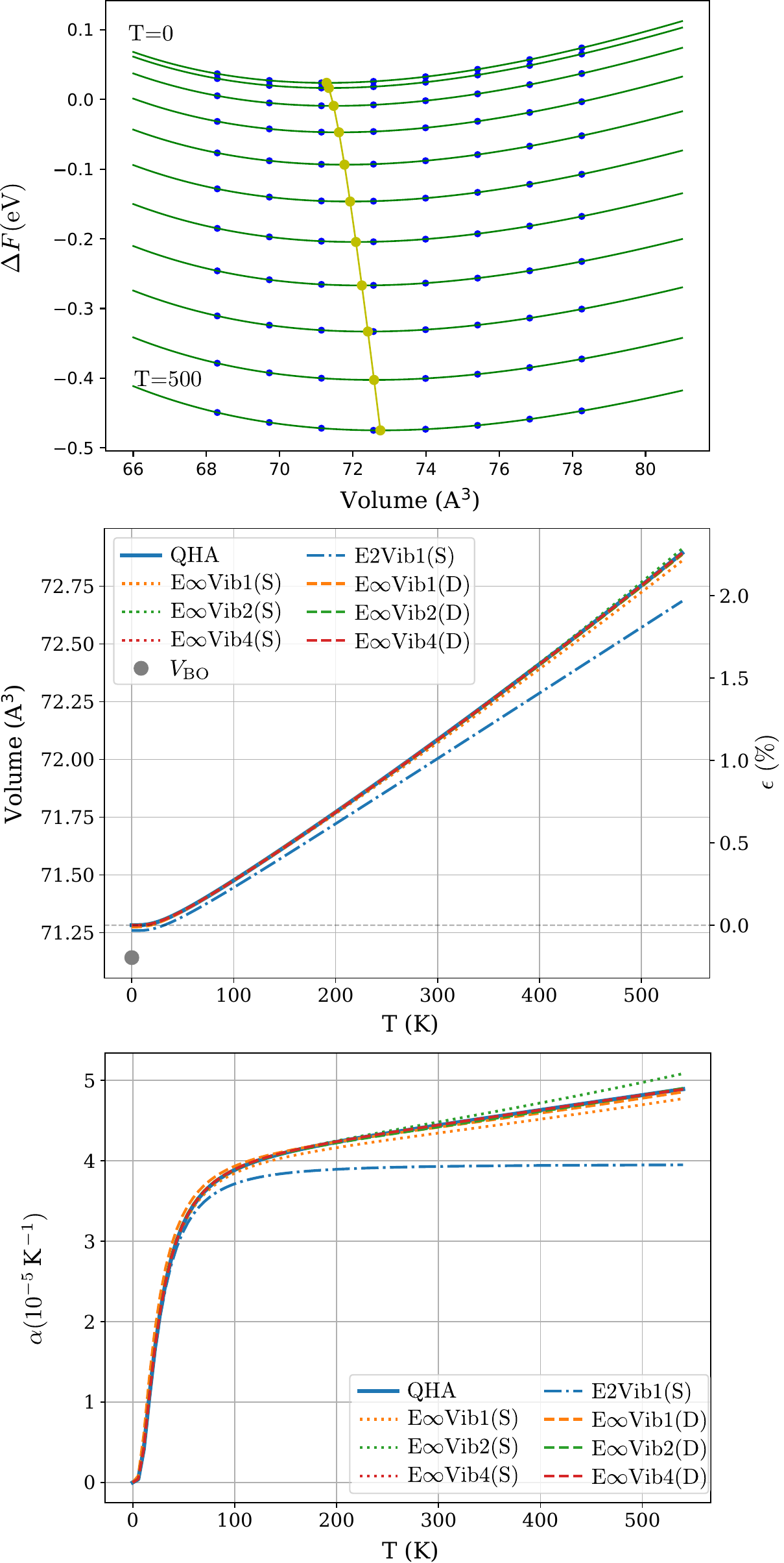}
		\caption{
  Detailed representation of the free energy, volume, and 
 thermal expansion coefficient of Bi, for a temperature range from 0 K to 550 K.
	Same conventions as in Fig.~\ref{fig:Si}.
	}
	\label{fig:Bi}
\end{figure}

\begin{figure}[!]	
	\includegraphics[width=0.45\textwidth]{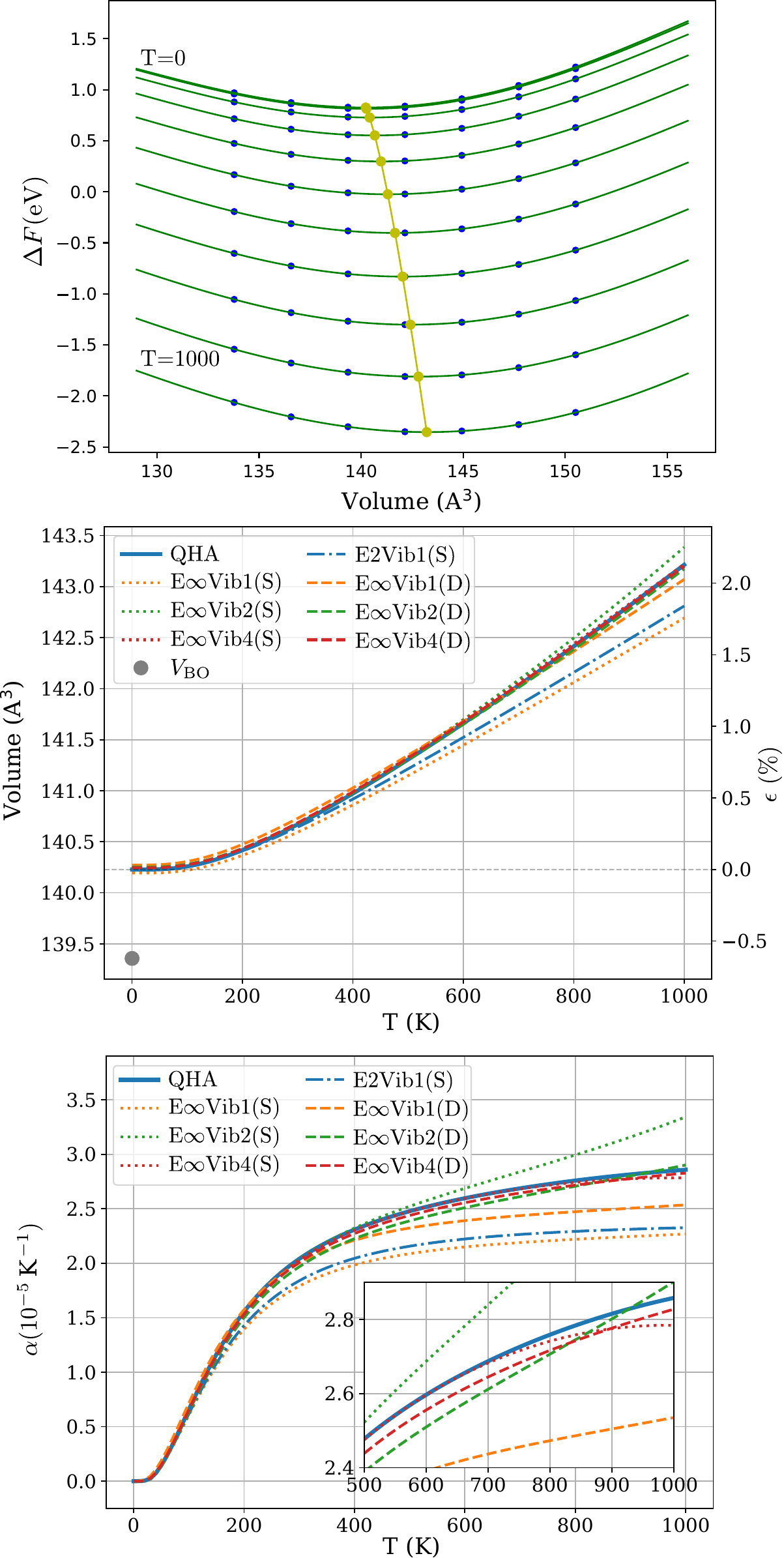}
		\caption{
    Detailed representation of the free energy, volume, and 
 thermal expansion coefficient of \ce{ZrO2}, for a temperature range from 0 K to 1000 K.
Same conventions as in Fig.~\ref{fig:Si}.
		The inset displays a zoomed-in view of the range $2.4\times10^{-5}$ to $2.9\times10^{-5}$ at temperatures 500 K to 1000 K.
	}
	\label{fig:ZrO2}
\end{figure}

In Figs.~\ref{fig:Si},~\ref{fig:Bi}, and ~\ref{fig:ZrO2}, the detailed representation of a temperature-dependent free energy, volume and thermal expansion coefficient is presented for Si, Bi,  and \ce{ZrO2}, illustrating different situations encountered when exploring the accuracy of the approximations that we study. 
The Supplemental Material~\cite{SupMat} (SM) provides similar information for the remaining materials.
The plots cover the temperature range from 0 K to 1000 K, except for Bi and Al, whose melting points are lower than 1000 K (approximately 545 and 933 K, respectively). 
All models are fitted to fourth-degree polynomials, as 
mentioned in Sec.~\ref{sec:IIF}.
Experimental data are not included in the figures, as the primary goal of this paper is to reproduce v-ZSISA-QHA results using cheaper approximations. 
To predict experimental results, a thorough comparison of DFT calculations across different functionals and pseudopotentials would be necessary. 
Additionally, anharmonicities in material properties may influence thermal expansion behavior, as discussed in the introduction section.

In these figures, the top panel shows free energies at different temperatures, against volumes in the QHA approach, for cubic materials, or v-ZSISA-QHA for other materials.
For the sake of brevity, ``QHA" is mentioned in the legend of the figures as well as in the text. 
Interpolated lines connect yellow points marking the minima of each free energy curve. The middle panel shows optimal volumes plotted against temperatures, with a gray dot denoting the minima of BO DFT total energy. Even at zero temperature, the free energy minimum volume differs from the BO DFT one, due to zero-point motion.

Volumetric thermal expansion $\epsilon (T)$, given by
\begin{align}
	\epsilon (T) = \frac{V(T)-V(T=0)}{V(T=0)},
\end{align}
is provided in the left side ticks. 
To avoid double plotting, we use $V(T=0)$ from QHA models in the plots. This approach ensures identical plots during rescaling, as using exact values of $V(T=0)$ from each model may introduce slight changes due to errors on ZPLE. 
However, this difference is negligible.

{The bottom panel presents the} thermal expansion coefficient $\alpha (T)$, defined as
\begin{align}
	\alpha (T) = \frac{1}{V(T_c)} \frac{dV(T)}{dT},
\end{align}
where $T_c$ = 293 K is the reference temperature. 
{ Alternatively, the thermal expansion coefficient might be based on the temperature-dependent volume $V(T)$ instead of the reference volume $V(T_c)$. Using $V(T_c)$  is straightforward and common in standard definitions and basic experiments\cite{Corruccini1961}.}
{ It is well-adapted to describe the expansion when the volume change is rather small. 
In contrast, using $V(T)$  provides greater accuracy for materials with non-linear expansion or presenting large volume changes. 
Here, we use the standard approach with $V(T_c)$ . Note that the biggest of volume changes among all our materials, occurs for Al, with a 5\% change from 293 K to 900 K.}

\begin{table*}[!]
	\caption{
		For  five distinct approximations, namely, QHA -taken as reference-, E2Vib1(D), E$\infty$Vib1(D), E$\infty$Vib2(D), and E$\infty$Vib4(D), and for the five cubic materials (MgO, Si, GaAs, Al, and Cu), this table details the zero-point lattice expansion (ZPLE),  volume change with respect to 0~K, $\epsilon$, volumetric thermal expansion coefficient, $\alpha$, {BO pressure, $P_{\textrm{BO}}$,} and bulk modulus, B. ZPLE and $\epsilon$ are given in percents, $\alpha$ in K$^{-1}$, and $P_{\textrm{BO}}$ and B in GPa. 
		$\epsilon$
		and $\alpha$ are given at room temperature (@293~K) as well as at high temperature (@800~K),
		$P_{\textrm{BO}}$ at 800~K for $V(800~K)$ and Bulk Modulus values at 293~K for $V(293~K)$
		. 
		On the left are relative errors 
		$\delta_{\textrm{rQHA}}$(ZPLE), 
		$\delta_{\textrm{rQHA}}(\epsilon)$, 
		$\delta_{\textrm{rQHA}}(\alpha)$,
		$\delta_{\textrm{rQHA}}(P_{\textrm{BO}})$
		and $\delta_{\textrm{rQHA}}(B)$ with respect to QHA, all in percents. For errors smaller than 0.1\% the indication $\approx$0.0 is used.}
	\begin{ruledtabular}
		\begin{tabular}{llllll|lrrrrr}  
			& QHA & E2Vib1 & E$\infty$Vib1 & E$\infty$Vib2 & E$\infty$Vib4  & Relative Error  & E2Vib1 & E$\infty$Vib1 & E$\infty$Vib2 & E$\infty$Vib4 \\
			\hline
			MgO:\\
			ZPLE(\%) &   1.206 &  1.191 &  1.192 &  1.207 &  1.206  &
			$\delta_{\textrm{rQHA}}$(ZPLE) &  -1.2 & -1.2&  $\approx$0.0 & $\approx$0.0\\
			$\epsilon$@293~K(\%) &  0.488&   0.443 &  0.501 &  0.489 &   0.488
			&$\delta_{\textrm{rQHA}}(\epsilon$@293~K)& 
			-9.3  &  2.8  &  0.1      &  $\approx$0.0  \\
			$\alpha$@293~K&3.52$\times 10^{-5}$&  3.18$\times 10^{-5}$&  3.56$\times 10^{-5}$&  3.53$\times 10^{-5}$&  3.52$\times 10^{-5}$
			&$\delta_{\textrm{rQHA}}(\alpha$@293~K)   &-9.8   &  1.1   &  0.1       &  $\approx$0.0\\
			$\epsilon$@800~K(\%) & 2.801 & 2.386 & 2.783 & 2.797 & 2.801
			&$\delta_{\textrm{rQHA}}(\epsilon$@800~K)&
			-14.8 &  -0.6     &  -0.1     &  $\approx$0.0  \\
			$\alpha$@800~K & 5.25$\times 10^{-5}$& 4.08$\times 10^{-5}$&  5.06$\times 10^{-5}$&  5.21$\times 10^{-5}$&  5.24$\times 10^{-5}$
			&$\delta_{\textrm{rQHA}}(\alpha$@800~K) &  -22.2  &  -3.5      &  -0.7      &  $\approx$0.0  \\  
			$P_{\textrm{BO}}$@800~K & 5.76 & 5.19 & 5.72 & 5.76 & 5.76 &$\delta_{\textrm{rQHA}}(P_{\textrm{BO}}$@800~K)& -9.8  & -0.7         &  $\approx$0.0 & $\approx$0.0 \\
			$B$@293~K & 148.07 & 160.33 & 147.19 & 147.99 & 148.04 & $\delta_{\textrm{rQHA}}(B$@293~K) & 8.3  & -0.6         & $\approx$0.0 & $\approx$0.0 \\
			Si:\\
			ZPLE(\%) &  0.480 &  0.476 &  0.475 &  0.481 &  0.480 &
			$\delta_{\textrm{rQHA}}$(ZPLE) &  -1.0 & -1.1&  $\approx$0.0 & $\approx$0.0 \\
			$\epsilon$@293~K(\%) &  0.050&   0.040 &  0.071 &  0.051 &   0.050
			&$\delta_{\textrm{rQHA}}(\epsilon$@293~K) &
			-20.3 &  40.7 &  0.8      &  $\approx$0.0  \\ 
			$\alpha$@293~K& 7.42$\times 10^{-6}$&  6.76$\times 10^{-6}$&  8.26$\times 10^{-6}$&  7.46$\times 10^{-6}$&  7.45$\times 10^{-6}$
			&$\delta_{\textrm{rQHA}}(\alpha$@293~K)  & -8.9   &  11.2  &  0.4       &  0.4    \\
			$\epsilon$@800~K(\%) &  0.606 & 0.531 & 0.645 & 0.604 & 0.606
			&$\delta_{\textrm{rQHA}}(\epsilon$@800~K)&
			-12.3 &  6.4      &  -0.2     &  $\approx$0.0  \\ 
			$\alpha$@800~K & 1.28$\times 10^{-5}$& 1.09$\times 10^{-5}$&  1.27$\times 10^{-5}$&  1.27$\times 10^{-5}$&  1.28$\times 10^{-5}$
			&$\delta_{\textrm{rQHA}}(\alpha$@800~K) & -14.6  &  -0.3      &  -0.3      &  0.1    \\  
			$P_{\textrm{BO}}$@800~K & 0.99 & 0.92 & 1.02 & 0.99 & 0.99 &$\delta_{\textrm{rQHA}}(P_{\textrm{BO}}$@800~K)& -7.1  & 3.0          &  -0.1         & $\approx$0.0 \\
			$B$@293~K & 90.19  & 93.86  & 91.24  & 90.37  & 90.19  & $\delta_{\textrm{rQHA}}(B$@293~K) & 4.1  & 1.2          & 0.2          & $\approx$0.0 \\
			GaAs:\\
			ZPLE(\%) &   0.377 &  0.381 &  0.370 &  0.376 &  0.376  &
			$\delta_{\textrm{rQHA}}$(ZPLE) &   1.1 & -1.8&  $\approx$0.0 & $\approx$0.0\\
			$\epsilon$@293~K(\%) & 0.332&   0.311 &  0.351 &  0.332 &   0.331
			&$\delta_{\textrm{rQHA}}(\epsilon$@293~K) & 
			-6.1  &  5.9  &  $\approx$0.0 &  $\approx$0.0  \\ 
			$\alpha$@293~K&    2.00$\times 10^{-5}$&  1.86$\times 10^{-5}$&  2.04$\times 10^{-5}$&  2.00$\times 10^{-5}$&  2.00$\times 10^{-5}$
			&$\delta_{\textrm{rQHA}}(\alpha$@293~K) &  -7.0   &  1.9   &  -0.2      &  $\approx$0.0 \\
			$\epsilon$@800~K(\%) &   1.492 & 1.333 & 1.501 & 1.490 & 1.492
			&$\delta_{\textrm{rQHA}}(\epsilon$@800~K) &
			-10.7 &  0.6      &  -0.1     &  $\approx$0.0  \\ 
			$\alpha$@800~K & 2.48$\times 10^{-5}$& 2.07$\times 10^{-5}$&  2.40$\times 10^{-5}$&  2.48$\times 10^{-5}$&  2.48$\times 10^{-5}$
			&$\delta_{\textrm{rQHA}}(\alpha$@800~K) &  -16.7  &  -3.3      &  -0.1      &  $\approx$0.0\\  
			$P_{\textrm{BO}}$@800~K & 1.23 & 1.13 & 1.23 & 1.23 & 1.23 &$\delta_{\textrm{rQHA}}($@800~K)& -7.8  & 0.1          &  -0.1         & $\approx$0.0 \\
			$B$@293~K & 66.04  & 69.57  & 66.74  & 66.12  & 66.04  & $\delta_{\textrm{rQHA}}(B$@293~K) & 5.3  & 1.0          & 0.1          & $\approx$0.0 \\
			 
			Al:\\
			ZPLE(\%) & 	  0.996 &  0.962 &  0.981 &  0.996 &  0.996  &
			$\delta_{\textrm{rQHA}}$(ZPLE) &  -3.4 & -1.5&  $\approx$0.0 & $\approx$0.0\\
			$\epsilon$@293~K(\%) &  1.171&   1.048 &  1.186 &  1.172 &   1.171
			&$\delta_{\textrm{rQHA}}(\epsilon$@293~K) &
			-10.5 &  1.2  &  $\approx$0.0 &  $\approx$0.0  \\ 
			$\alpha$@293~K&   6.48$\times 10^{-5}$&  5.60$\times 10^{-5}$&  6.43$\times 10^{-5}$&  6.48$\times 10^{-5}$&  6.48$\times 10^{-5}$
			&$\delta_{\textrm{rQHA}}(\alpha$@293~K)  & -13.7  &  -0.8  &  $\approx$0.0  &  $\approx$0.0 \\
			$\epsilon$@800~K(\%) &	  5.255 & 4.095 & 5.078 & 5.234 & 5.254
			&$\delta_{\textrm{rQHA}}(\epsilon$@800~K)&
			-22.1 &  -3.4     &  -0.4     &  $\approx$0.0  \\ 
			$\alpha$@800~K & 9.54$\times 10^{-5}$& 6.08$\times 10^{-5}$&  8.68$\times 10^{-5}$&  9.39$\times 10^{-5}$&  9.52$\times 10^{-5}$ 
			&$\delta_{\textrm{rQHA}}(\alpha$@800~K) &  -36.2  &  -9.0      &  -1.5      &  -0.2   \\
			$P_{\textrm{BO}}$@800~K & 4.39 & 3.67 & 4.28 & 4.38 & 4.39 &$\delta_{\textrm{rQHA}}(P_{\textrm{BO}}$@800~K)& -16.6 & -2.5         &  -0.3         & $\approx$0.0 \\
			$B$@293~K & 74.72  & 84.79  & 75.20  & 74.72  & 74.71  & $\delta_{\textrm{rQHA}}(B$@293~K) & 13.5 & 0.6          & $\approx$0.0 & $\approx$0.0 \\  
				Cu:\\
			ZPLE(\%) &    0.580 &  0.575 &  0.566 &  0.580 &  0.580    &
			$\delta_{\textrm{rQHA}}$(ZPLE) &  -1.0 & -2.5&  $\approx$0.0 & $\approx$0.0 \\
			$\epsilon$@293~K(\%) &     0.856&   0.799 &  0.871 &  0.856 &   0.856
			&$\delta_{\textrm{rQHA}}(\epsilon$@293~K)& 
			-6.6  &  1.8  &  $\approx$0.0 &  $\approx$0.0  \\ 
			$\alpha$@293~K&  4.44$\times 10^{-5}$&  4.04$\times 10^{-5}$&  4.46$\times 10^{-5}$&  4.45$\times 10^{-5}$&  4.45$\times 10^{-5}$
			&$\delta_{\textrm{rQHA}}(\alpha$@293~K) &-9.1   &  0.3   &  $\approx$0.0  &  $\approx$0.0\\
			$\epsilon$@800~K(\%) &  3.455 & 2.959 & 3.438 & 3.452 & 3.455
			&$\delta_{\textrm{rQHA}}(\epsilon$@800~K) & 
			-14.4 &  -0.5     &  $\approx$0.0 &  $\approx$0.0  \\ 
			$\alpha$@800~K &  5.68$\times 10^{-5}$& 4.29$\times 10^{-5}$&  5.52$\times 10^{-5}$&  5.65$\times 10^{-5}$&  5.68$\times 10^{-5}$
			&$\delta_{\textrm{rQHA}}(\alpha$@800~K)  & -24.4  &  -2.9      &  -0.5      &  $\approx$0.0  \\ 
			$P_{\textrm{BO}}$@800~K & 6.17 & 5.48 & 6.13 & 6.16 & 6.17 &$\delta_{\textrm{rQHA}}(P_{\textrm{BO}}$@800~K)& -11.2 & -0.7         &  $\approx$0.0 & $\approx$0.0 \\
			$B$@293~K & 159.48 & 173.98 & 159.61 & 159.40 & 159.47 & $\delta_{\textrm{rQHA}}(B$@293~K) & 9.1  & $\approx$0.0 & $\approx$0.0 & $\approx$0.0 
			\label{tab:resultsIII}	
		\end{tabular}
	\end{ruledtabular}
\end{table*}
\begin{table*}[!]
	\caption{	
		For  five distinct approximations, namely QHA -taken as reference-, E2Vib1(D), E$\infty$Vib1(D), E$\infty$Vib2(D), and E$\infty$Vib4(D), and for the hexagonal and orthorhombic materials (ZnO, AlN, GaN, and YAlO$_3$), this table details the zero-point lattice expansion (ZPLE),  volume change with respect to 0 K, $\epsilon$, volumetric thermal expansion coefficient, $\alpha$, {BO pressure, $P_{\textrm{BO}}$,} and bulk modulus, B. ZPLE and $\epsilon$ are given in percents, $\alpha$ in K$^{-1}$, and $P_{\textrm{BO}}$  and B in GPa. 
		$\epsilon$
		and $\alpha$ are given at room temperature (293 K) as well as at high temperature (800 K), $P_{\textrm{BO}}$ at 800~K for $V(800~K)$ and Bulk Modulus values at 293~K for $V(293~K)$. For Bi, the high-temperature $\epsilon$
		and $\alpha$ are reported at 500 K instead of 800 K, because of its low melting point.
		On the left are relative errors
		$\delta_{\textrm{rQHA}}$(ZPLE), 
		$\delta_{\textrm{rQHA}}(\epsilon)$, 
		$\delta_{\textrm{rQHA}}(\alpha)$,
		$\delta_{\textrm{rQHA}}(P_{\textrm{BO}})$
		and $\delta_{\textrm{rQHA}}(B)$ 
		with respect to QHA,
		all in percents. For errors smaller than 0.1\% the indication $\approx$0.0 is used.}
	\begin{ruledtabular}
		\begin{tabular}{llllll|lrrrrr}	
			& QHA & E2Vib1 & E$\infty$Vib1 & E$\infty$Vib2 & E$\infty$Vib4  & Relative Error & E2Vib1 & E$\infty$Vib1 & E$\infty$Vib2 & E$\infty$Vib4 \\
			\hline
			ZnO:\\
			ZPLE(\%) &    0.719 &  0.711 &  0.709 &  0.719 &  0.719    &
			$\delta_{\textrm{rQHA}}$(ZPLE) &  -1.1 & -1.4&  $\approx$0.0 & $\approx$0.0 \\
			$\epsilon$@293~K(\%) &  0.109&   0.081 &  0.140 &  0.109 &   0.109
			&$\delta_{\textrm{rQHA}}(\epsilon$@293~K) & 
			-25.5 &  28.3 &  -0.3     &  -0.3      \\ 
			$\alpha$@293~K& 1.29$\times 10^{-5}$&  1.11$\times 10^{-5}$&  1.39$\times 10^{-5}$&  1.29$\times 10^{-5}$&  1.29$\times 10^{-5}$
			&$\delta_{\textrm{rQHA}}(\alpha$@293~K) &-13.8  &  8.2   &  $\approx$0.0  &  0.2    \\
			$\epsilon$@800~K(\%) & 	 1.065 & 0.860 & 1.086 & 1.061 & 1.065
			&$\delta_{\textrm{rQHA}}(\epsilon$@800~K) &
			-19.3 &  2.0      &  -0.4     &  $\approx$0.0  \\
			$\alpha$@800~K &2.24$\times 10^{-5}$& 1.71$\times 10^{-5}$&  2.10$\times 10^{-5}$&  2.24$\times 10^{-5}$&  2.25$\times 10^{-5}$    
			&$\delta_{\textrm{rQHA}}(\alpha$@800~K)&  -23.6  &  -6.4      &  -0.2      &  0.1    \\        
			$P_{\textrm{BO}}$@800~K & 2.49 & 2.20 & 2.50 & 2.48 & 2.49 &$\delta_{\textrm{rQHA}}(P_{\textrm{BO}}$@800~K)& -11.5 & 0.6          &  -0.2         & $\approx$0.0 \\
			$B$@293~K & 137.72 & 146.96 & 140.37 & 137.96 & 137.69 & $\delta_{\textrm{rQHA}}(B$@293~K) & 6.7  & 1.9          & 0.2          & $\approx$0.0 \\
				
			AlN:\\
			ZPLE(\%) &  0.838 &  0.824 &  0.825 &  0.838 &  0.838   &
			$\delta_{\textrm{rQHA}}$(ZPLE) &  -1.7 & -1.6&  $\approx$0.0 & $\approx$0.0\\
			$\epsilon$@293~K(\%) &  0.092&   0.085 &  0.098 &  0.092 &   0.092
			&$\delta_{\textrm{rQHA}}(\epsilon$@293~K) & 
			-7.3  &  6.1  &  $\approx$0.0 &  $\approx$0.0  \\ 
			$\alpha$@293~K&   9.27$\times 10^{-6}$&  8.69$\times 10^{-6}$&  9.61$\times 10^{-6}$&  9.27$\times 10^{-6}$&  9.27$\times 10^{-6}$
			&$\delta_{\textrm{rQHA}}(\alpha$@293~K) &  -6.3   &  3.7   &  $\approx$0.0  &  $\approx$0.0 \\
			$\epsilon$@800~K(\%) &   0.846 & 0.783 & 0.861 & 0.846 & 0.846
			&$\delta_{\textrm{rQHA}}(\epsilon$@800~K) &
			-7.5  &  1.7      &  $\approx$0.0 &  $\approx$0.0  \\ 
			$\alpha$@800~K & 1.78$\times 10^{-5}$& 1.62$\times 10^{-5}$&  1.78$\times 10^{-5}$&  1.78$\times 10^{-5}$&  1.78$\times 10^{-5}$ 
			&$\delta_{\textrm{rQHA}}(\alpha$@800~K)&  -9.4   &  $\approx$0.0  &  $\approx$0.0  &  $\approx$0.0 \\  
			$P_{\textrm{BO}}$@800~K & 3.25 & 3.10 & 3.25 & 3.25 & 3.25 &$\delta_{\textrm{rQHA}}($@800~K)& -4.5  & $\approx$0.0 &  $\approx$0.0 & $\approx$0.0 \\ 
			$B$@293~K & 195.15 & 202.64 & 193.75 & 195.10 & 195.15 & $\delta_{\textrm{rQHA}}(B$@293~K) & 3.8  & -0.7         & $\approx$0.0 & $\approx$0.0 \\
				
			GaN:\\
			ZPLE(\%) & 0.724 &  0.700 &  0.711 &  0.724 &  0.724   &
			$\delta_{\textrm{rQHA}}$(ZPLE)&  -3.3 & -1.8&  $\approx$0.0 & $\approx$0.0 \\
			$\epsilon$@293~K(\%) &    0.152&   0.142 &  0.161 &  0.152 &   0.152
			&$\delta_{\textrm{rQHA}}(\epsilon$@293~K) & 
			-7.0  &  5.9  &  $\approx$0.0 &  $\approx$0.0  \\ 
			$\alpha$@293~K&  1.21$\times 10^{-5}$&  1.14$\times 10^{-5}$&  1.25$\times 10^{-5}$&  1.21$\times 10^{-5}$&  1.21$\times 10^{-5}$
			&$\delta_{\textrm{rQHA}}(\alpha$@293~K)&  -6.5   &  3.3   &  $\approx$0.0  &  $\approx$0.0 \\
			$\epsilon$@800~K(\%) & 	 1.018 & 0.931 & 1.033 & 1.020 & 1.018
			&$\delta_{\textrm{rQHA}}(\epsilon$@800~K) & 
			-8.6  &  1.5      &  $\approx$0.0 &  $\approx$0.0  \\ 
			$\alpha$@800~K &  1.97$\times 10^{-5}$& 1.75$\times 10^{-5}$&  1.96$\times 10^{-5}$&  1.97$\times 10^{-5}$&  1.97$\times 10^{-5}$ 
			&$\delta_{\textrm{rQHA}}(\alpha$@800~K)  &  -11.4  &  -0.6      &  $\approx$0.0  &  $\approx$0.0 \\   
			$P_{\textrm{BO}}$@800~K & 3.11 & 2.92 & 3.12 & 3.11 & 3.11 &$\delta_{\textrm{rQHA}}(P_{\textrm{BO}}$@800~K)& -6.2  & 0.1          &  $\approx$0.0 & $\approx$0.0 \\
			$B$@293~K & 182.04 & 190.04 & 181.35 & 182.01 & 182.04 & $\delta_{\textrm{rQHA}}(B$@293~K) & 4.4  & -0.4         & $\approx$0.0 & $\approx$0.0 \\
				
			\ce{YAlO3}:\\
			ZPLE(\%) &   0.865 &  0.857 &  0.852 &  0.865 &  0.865    &
			$\delta_{\textrm{rQHA}}$(ZPLE)&  -1.0 & -1.5&  $\approx$0.0 & $\approx$0.0\\
			$\epsilon$@293~K(\%) &	 0.311&   0.288 &  0.324 &  0.311 &   0.311
			&$\delta_{\textrm{rQHA}}(\epsilon$@293~K) &
			-7.5  &  4.1  &  $\approx$0.0 &  $\approx$0.0  \\ 
			$\alpha$@293~K& 2.17$\times 10^{-5}$&  1.99$\times 10^{-5}$&  2.21$\times 10^{-5}$&  2.17$\times 10^{-5}$&  2.17$\times 10^{-5}$
			&$\delta_{\textrm{rQHA}}(\alpha$@293~K)  & -8.2   &  2.1   &  $\approx$0.0  &  $\approx$0.0 \\
			$\epsilon$@800~K(\%) & 1.751 & 1.545 & 1.750 & 1.751 & 1.751
			&
			$\delta_{\textrm{rQHA}}(\epsilon$@800~K) & 
			-11.7 &  $\approx$0.0 &  $\approx$0.0 &  $\approx$0.0  \\ 
			$\alpha$@800~K &3.24$\times 10^{-5}$& 2.68$\times 10^{-5}$&  3.14$\times 10^{-5}$&  3.24$\times 10^{-5}$&  3.24$\times 10^{-5}$  
			&$\delta_{\textrm{rQHA}}(\alpha$@800~K)&  -17.2  &  -3.2      &  $\approx$0.0  &  $\approx$0.0 \\ 
			$P_{\textrm{BO}}$@800~K & 5.01 & 4.62 & 4.99 & 5.01 & 5.01 &$\delta_{\textrm{rQHA}}(P_{\textrm{BO}}$@800~K)& -7.7  & -0.4         &  $\approx$0.0 & $\approx$0.0 \\
			$B$@293~K & 194.07 & 205.71 & 193.86 & 193.93 & 194.07 & $\delta_{\textrm{rQHA}}(B$@293~K) & 6.0  & -0.1         & $\approx$0.0 & $\approx$0.0 
			\label{tab:resultsIV}
		\end{tabular}
	\end{ruledtabular}
\end{table*}
\begin{table*}[!]
	\caption{	
		For  five distinct approximations, namely QHA -taken as reference-, E2Vib1(D), E$\infty$Vib1(D), E$\infty$Vib2(D), and E$\infty$Vib4(D), and for the rhombohedral and monoclinic materials (Bi, CaCO3, and ZrO$_2$)
		, this table details the zero-point lattice expansion (ZPLE),  volume change with respect to 0 K, $\epsilon$, volumetric thermal expansion coefficient, $\alpha$, BO pressure, $P_{\textrm{BO}}$, and bulk modulus, B. ZPLE and $\epsilon$ are given in percents, $\alpha$ in K$^{-1}$, and $P_{\textrm{BO}}$ and B in GPa. 
		$\epsilon$
		and $\alpha$ are given at room temperature (293 K) as well as at high temperature (800 K), $P_{\textrm{BO}}$ at 800~K for $V(800~K)$ and Bulk Modulus values at 293~K for $V(293~K)$. For Bi, the high-temperature $\epsilon$
		and $\alpha$ are reported at 500 K instead of 800 K, because of its low melting point.
		On the left are relative errors
		$\delta_{\textrm{rQHA}}$(ZPLE), 
		$\delta_{\textrm{rQHA}}(\epsilon)$, 
		$\delta_{\textrm{rQHA}}(\alpha)$,
		$\delta_{\textrm{rQHA}}(P_{\textrm{BO}})$
		and $\delta_{\textrm{rQHA}}(B)$ 
		with respect to QHA,
		all in percents. For errors smaller than 0.1\% the indication $\approx$0.0 is used.}
	\begin{ruledtabular}
		\begin{tabular}{llllll|lrrrrr}	
			& QHA & E2Vib1 & E$\infty$Vib1 & E$\infty$Vib2 & E$\infty$Vib4  & Relative Error & E2Vib1 & E$\infty$Vib1 & E$\infty$Vib2 & E$\infty$Vib4 \\
			\hline
			
			Bi:\\
			ZPLE(\%) &   0.198 &  0.164 &  0.187 &  0.198 &  0.198    &
			$\delta_{\textrm{rQHA}}$(ZPLE) & -17.1 & -5.5&  $\approx$0.0 & $\approx$0.0 \\
			$\epsilon$@293~K(\%) &    1.094&   1.016 &  1.104 &  1.091 &   1.093
			&$\delta_{\textrm{rQHA}}(\epsilon$@293~K) & 
			-7.1  &  1.0  &  -0.3     &  $\approx$0.0  \\ 
			$\alpha$@293~K&   4.42$\times 10^{-5}$&  3.93$\times 10^{-5}$&  4.40$\times 10^{-5}$&  4.41$\times 10^{-5}$&  4.43$\times 10^{-5}$
			&$\delta_{\textrm{rQHA}}(\alpha$@293~K)& -11.1  &  -0.4  &  -0.3      &  0.2   \\
			$\epsilon$@500K(\%) &  2.062 & 1.840 & 2.064 & 2.055 & 2.061
			&$\delta_{\textrm{rQHA}}(\epsilon$@500K) &
			-10.7 &  0.1      &  -0.3     &  $\approx$0.0  \\
			$\alpha$@500K &  4.81$\times  10^{-5}$&  3.95$\times  10^{-5}$&  4.78$\times  10^{-5}$&  4.81$\times  10^{-5}$&  4.81$\times  10^{-5}$
			&$\delta_{\textrm{rQHA}}(\alpha$@500K) & -17.9  &  -0.7      &  $\approx$0.0  &  0.1     \\
			$P_{\textrm{BO}}$@800~K & 0.62 & 0.55 & 0.62 & 0.62 & 0.62 &$\delta_{\textrm{rQHA}}(P_{\textrm{BO}}$@800~K)& -10.6 & -0.3         &  -0.2         & $\approx$0.0 \\
			$B$@293~K & 27.64  & 30.74  & 27.84  & 27.75  & 27.65  & $\delta_{\textrm{rQHA}}(B$@293~K) & 11.2 & 0.7          & 0.4          & $\approx$0.0 \\
				
			\ce{CaCO3}:\\
			ZPLE(\%) &   0.864 &  0.851 &  0.853 &  0.864 &  0.864  &
			$\delta_{\textrm{rQHA}}$(ZPLE) &  -1.5 & -1.2&  $\approx$0.0 & $\approx$0.0 \\
			$\epsilon$@293~K(\%) & 	0.282&   0.212 &  0.321 &  0.281 &   0.282
			&$\delta_{\textrm{rQHA}}(\epsilon$@293~K) & 
			-24.8 &  13.9 &  -0.4     &  $\approx$0.0  \\
			$\alpha$@293~K&  2.18$\times 10^{-5}$&  1.76$\times 10^{-5}$&  2.26$\times 10^{-5}$&  2.18$\times 10^{-5}$&  2.20$\times 10^{-5}$
			&$\delta_{\textrm{rQHA}}(\alpha$@293~K)  & -19.3  &  3.4   &  $\approx$0.0  &  0.5     \\
			$\epsilon$@800~K(\%) &  1.782 & 1.288 & 1.703 & 1.776 & 1.781
			&$\delta_{\textrm{rQHA}}(\epsilon$@800~K) & 
			-27.8 &  -4.4     &  -0.3     &  $\approx$0.0  \\ 
			$\alpha$@800~K &  3.60$\times 10^{-5}$& 2.30$\times 10^{-5}$&  3.03$\times 10^{-5}$&  3.62$\times 10^{-5}$&  3.60$\times 10^{-5}$
			&$\delta_{\textrm{rQHA}}(\alpha$@800~K)&  -36.1  &  -15.9     &  0.6       &  $\approx$0.0   \\
			$P_{\textrm{BO}}$@800~K & 1.90 & 1.56 & 1.84 & 1.90 & 1.90 &$\delta_{\textrm{rQHA}}(P_{\textrm{BO}}$@800~K)& -18.1 & -3.1         &  -0.2         & $\approx$0.0 \\
			$B$@293~K & 70.22  & 77.67  & 72.95  & 70.48  & 70.24  & $\delta_{\textrm{rQHA}}(B$@293~K) & 10.6 & 3.9          & 0.4          & $\approx$0.0 \\

			\ce{ZrO2}:\\
			ZPLE(\%) &  0.625 &  0.628 &  0.657 &  0.641 &  0.637 &
			$\delta_{\textrm{rQHA}}$(ZPLE) &   0.5 &  5.1&  2.6      & 1.8 \\
			$\epsilon$@293~K(\%) &  0.301&   0.280 &  0.312 &  0.294 &   0.301
			&$\delta_{\textrm{rQHA}}(\epsilon$@293~K) &
			-7.0  &  3.6  &  -2.3     &  -0.2      \\ 
			$\alpha$@293~K&  2.02$\times 10^{-5}$&  1.82$\times 10^{-5}$&  1.98$\times 10^{-5}$&  1.94$\times 10^{-5}$&  1.98$\times 10^{-5}$
			&$\delta_{\textrm{rQHA}}(\alpha$@293~K)   & -9.7   &  -1.7  &  -3.5      &  -1.6   \\
			$\epsilon$@800~K(\%) &   1.559 & 1.374 & 1.492 & 1.521 & 1.548
			&$\delta_{\textrm{rQHA}}(\epsilon$@800~K) & 
			-11.9 &  -4.3     &  -2.4     &  -0.7      \\ 
			$\alpha$@800~K &  2.76$\times 10^{-5}$& 2.29$\times 10^{-5}$&  2.47$\times 10^{-5}$&  2.71$\times 10^{-5}$&  2.72$\times 10^{-5}$	
			&$\delta_{\textrm{rQHA}}(\alpha$@800~K)  & -16.9  &  -10.4     &  -1.9      &  -1.5    \\       
			$P_{\textrm{BO}}$@800~K & 3.93 & 3.60 & 3.86 & 3.89 & 3.93 &$\delta_{\textrm{rQHA}}(P_{\textrm{BO}}$@800~K)& -8.4  & -1.7         &  -1.0         & $\approx$0.0 \\
			$B$@293~K & 170.30 & 174.44 & 179.68 & 173.91 & 170.56 & $\delta_{\textrm{rQHA}}(B$@293~K) & 2.4  & 5.5          & 2.1          & 0.2         
			\label{tab:resultsV}
		\end{tabular}
	\end{ruledtabular}
\end{table*}

Tables ~\ref{tab:resultsIII} ,\ref{tab:resultsIV}, and\ref{tab:resultsV}
list Zero Point Lattice Expansion (ZPLE), Volumetric Thermal Expansion ($\epsilon(T)$), Volumetric Thermal Expansion Coefficient ($\alpha(T)$), Born-Oppenheime pressure ($P_{\textrm{BO}}$), and bulk modulus (B) derived from five different approximations: QHA, E2Vib1(S), E$\infty$Vib1(D), E$\infty$Vib2(D), and E$\infty$Vib4(D) for our twelve materials. 
The table provides detailed values for Volumetric Thermal Expansion and Volumetric Thermal Expansion Coefficient, measured in K$^{-1}$, at temperatures of 293 K and 800 K (with a specific condition for Bi at 500 K). Additionally, $P_{\textrm{BO}}$ and Bulk Modulus values at  the optimized volume $V(T)$ are presented in GPa for 800~K and 293~K, respectively.
$P_{\textrm{BO}}$ is determined by computing the derivative of $E_{\textrm{BO}}$, fitted to a fourth-order polynomial across all volumes, with respect to volume, at the optimized volume  $V(T)$ in each model.
Furthermore, the table offers insights into the respective 
relative errors associated with each model when compared to QHA,
which forms the basis of a quantitative assessment of the merit of each 
approximation, as follows.

For all models, one observes rather accurate zero-point lattice expansion (ZPLE), that agrees with the QHA model within 5\%, with the exception of Bi. 
As temperature increases, the accuracy of the volume change
predictions deviates across models.
For E2Vib1, across different properties, deviations are often more than 10\%. There is an improvement from E2Vib1 to  E$\infty$Vib1(S) (not mentioned in the table), and further to 
E$\infty$Vib1(D), although several
cases exhibit also deviations
on the order of 10\% in the latter case. 
All other approximations show agreement with QHA within a couple of percent, or even much lower.

We focus now on the three example materials.
For Si, the thermal expansion is less than 0.9\% in the target temperature range.
The relative error in the volume change at 293 K from 
E2Vib1,  E$\infty$Vib1(S) (not mentioned in the table), and 
E$\infty$Vib1(D) might seem deceptive being in the dozens of percent range.
However, 
at low temperatures, negative thermal expansion is observed,
and the volume change at 293 K is
small. This affects the denominator of the relative difference
of volume changes.
Anyhow, models based on E$\infty$Vib2 and E$\infty$Vib4 with both symmetric and displaced selection exhibit strong agreement with the QHA model, all properties
being within 1\% of the reference QHA.
The results emphasize that accurate thermal expansion properties can be obtained with phonon calculations in three volumes, consistently aligning with the QHA model.

A numerical deviation within 1\% of the reference QHA one is also observed for E$\infty$Vib2 and E$\infty$Vib4 with both symmetric and displaced selections, for MgO, GaAs, Al (with one exception at 1.5\%), Cu, ZnO, AlN, GaN, YAlO$_3$, Bi, and CaCO$_3$.
 
Bismuth, exhibiting metallic behavior, is characterized as a remarkably soft metal. 
In the context of Fig.~\ref{fig:Bi}, its free energy profile resembles a wide parabola, enabling the determination of a bulk modulus of 27.5 GPa through fitting to the Vinet equation of state (in decent agreement with the experimental value of 31 GPa). 
Considering the melting point of Bi at 544 K, the analysis covers the temperature range of 0 to 540 K. The calculated volumetric thermal expansion stands at 2\% at 500 K.
Comparing thermal expansion coefficients, models utilizing displaced selections (D) and E$\infty$Vib4(S) exhibit excellent alignment with QHA, as already mentioned. 

More interestingly, E$\infty$Vib1(S) and 
E$\infty$Vib1(D) also do very well, with the exception of the ZPLE. For all the other properties, a deviation by less than 1\% is observed for E$\infty$Vib1(D).
However, Bi is the only example of such excellent behaviour of
E$\infty$Vib1(S) and 
E$\infty$Vib1(D). 
All the other cases deliver some errors for 
E$\infty$Vib1(D)
that are bigger than 2.9\% at least, and rather often in the 1-10\% range.

The situation for Zirconia (\ce{ZrO2}), the most complex material in our study, with 9 internal degrees of freedom and a low-symmetry space group of $P2_1/c$, is less satisfactory. 
As depicted in Fig.~\ref{fig:ZrO2}, and in Table~\ref{tab:resultsV},
even the Vib2 models encounter limitations at high temperatures.
Focusing on E$\infty$Vib2(D),
most of the relative errors are in the range between one and four percents. Of course, this might be considered enough depending on the purpose of the calculations. 
Still, it is clearly bigger than for the other materials.
For those materials, the vibrational free energies cannot be accurately described solely by approximating the Taylor series to the second derivative.
E$\infty$Vib2(D) is doing better, keeping within 2\%.

The E$\infty$Vib4(S) model closely mirrors QHA behavior below 700 K. Five volume points are required for an accurate description of thermal properties in this case.

For the remaining materials presented in the Supplemental Material, both Vib2 and Vib4 models exhibit good alignment with QHA. However, in \ce{CaCO3}, E$\infty$Vib2(S) fails to accurately represent the system.

Regarding \ce{YAlO3}, featuring a primitive cell with 20 atoms and 7 internal degrees of freedom, the results indicate excellent agreement of Vib2 and Vib4 models with QHA. Interestingly, even E$\infty$Vib1(D) performs well below 600 K, aligning closely with QHA. Given the expensive computational cost of phonon calculations for this material, the ability to compute thermal properties with a minimal number of phonon calculations, such as three for Vib2, proves highly valuable.

The E$\infty$Vib2(D) is thus an excellent approximation, providing for nearly all properties and materials an accuracy with respect to the QHA better than one percent, with two exceptions, the thermal expansion of Al at 800 K, obtained within 1.5\% of the reference, and the different properties of ZrO$_2$, obtained within 3.5\%. 
Whether this proves adequate for the purpose of the intended usage is left to the appreciation of the reader.

\section{CONCLUSION}
\label{sec:conclusion}

In this study, we investigated the accuracy of various approximations for determining volumetric thermal expansion using first-principles methods, taking the Quasi-Harmonic Approximation with 
volume-constrained Zero Static Internal Stress Approximation as a reference. While the v-ZSISA-QHA method is widely accepted, its computational demands due to numerous phonon spectra calculations can be substantial. 
Some of our approximations significantly reduce this computational burden while maintaining precise outcomes for volumetric thermal expansion determination.

Through extensive analysis across a representative set of twelve materials, encompassing various crystallographic systems from simple FCC structures to more complex hexagonal, orthorhombic, rhombohedral, and monoclinic systems, we determined the minimal number of phonon spectra calculations needed to achieve precise results. We found that for most materials, three full phonon spectra calculations, corresponding to quadratic order, are sufficient to determine thermal expansion with excellent (less than 1\% error) to reasonable (a couple of percents) accuracy, and near-perfect agreement with the v-ZSISA-QHA method is achieved with five phonon spectra.

Our results underscore the trade-off between computational efficiency and accuracy in first-principles calculations of volumetric thermal expansion. While simpler approximations reduce computational demands, they may sacrifice accuracy, particularly at higher temperatures or for complex materials. Nevertheless, our findings demonstrate that with careful selection of approximations and appropriate computational strategies, accurate predictions of thermal expansion properties can be obtained with reduced computational costs.

This study lays the groundwork for future research aimed at multi-dimensional generalizations beyond volumetric thermal expansion, with the potential for even greater computational efficiency and accuracy. By refining computational models and exploring new materials, we can advance our understanding of thermal expansion behavior and facilitate the design of materials for various applications.

{The results presented here might also be useful as reference results to gauge the accuracy of machine learning (ML) methods. 
Indeed, these allow one to drastically decrease the time needed for the evaluation of the Born-Oppenheimer energy surface, as well as the calculation of phonon spectra.
On one hand, some ML methods even allow automatic differentiation, like PyTorch or JAX. These methods can indeed make obtaining higher-order derivatives and Taylor series expansions more efficient. 
On the other hand, with ML methods, QHA with a large number of volumes can be performed very quickly, making the approximations studied in the present work less critical. 
However, our work aims to provide approximations that are especially useful for Density Functional Theory (DFT) and Density Functional Perturbation Theory (DFPT) calculations, with high computational cost, with one possible usage of these as a reference to test ML methods.} 

\begin{acknowledgments}
We acknowledge discussions with Dr. Matteo Giantomassi, concerning the implementation. This work has been supported by the Fonds de la Recherche
Scientifique (FRS-FNRS Belgium) through the PdR Grant No.
T.0103.19 – ALPS. 
It is an outcome of the Shapeable
2D magnetoelectronics by design project (SHAPEme, EOS
Project No. 560400077525) that has received funding from
the FWO and FRS-FNRS under the Belgian Excellence of
Science (EOS) program.
Computational resources have been provided by the supercomputing facilities of the Universit\'e catholique de Louvain (CISM/UCL) and the Consortium des Equipements de Calcul Intensif en F\'ed\'eration Wallonie Bruxelles (CECI) funded by the FRS-FNRS under Grant No. 2.5020.11.
\end{acknowledgments}

\bibliography{ThermalExpansion}

\clearpage
\onecolumngrid
\renewcommand{\theequation}{S\arabic{equation}}
\renewcommand{\thesection}{S\arabic{section}}
\renewcommand{\thesubsection}{S\arabic{section}.\arabic{subsection}}
\renewcommand{\thefigure}{S\arabic{figure}}
\renewcommand{\thetable}{S\arabic{table}}
\renewcommand{\thepage}{S\arabic{page}}
\setcounter{section}{0}
\setcounter{figure}{0}
\setcounter{table}{0}
\setcounter{page}{1}
\setcounter{equation}{0}

	\begin{center}
			{\Large \textbf{Supplemental Material: Approximations in first-principle volumic thermal expansion determination}}\\[10pt]
		 Samare Rostami\textsuperscript{1,*} and Xavier Gonze\textsuperscript{1} \\[5pt]
		\textsuperscript{1}\textit {European Theoretical Spectroscopy Facility, Institute of Condensed Matter and Nanosciences,\\ Universit\'{e} catholique de Louvain, Chemin des \'{e}toiles 8,\\ bte L07.03.01, B-1348 Louvain-la-Neuve, Belgium }
	\end{center}
	
	~\\~\\
		In the Supplemental Material, the formulas used to compute the numerical derivatives of the free energy, as well as the free energies as a function of volume, the volume as a function of temperature, and thermal expansion coefficient as a function of temperature, are represented for the different materials (except Si, Bi  and ZrO$_2$, whose similar figures are shown in the body of the paper). For non-cubic materials, the v-ZSISA-QHA is used as a reference, albeit abbreviated as QHA in the legends. Additionally, the optimized structures of   ZrO$_2$ and YAlO$_3$ are  included.
  
%
\section{ }
Here, we outline the formulas used to compute the numerical derivatives of the free energy needed for the different approximations. The computation of the numerical derivative of the entropy proceeds likewise.\\

The equation of free energies are rewritten in terms of $V$, $ V^{\bullet}$ and $\Delta^{\bullet} V = V-V^{\bullet}$ as follows:
\begin{align}
F_{\textrm{E}\infty \textrm{Vib}1}(V,T) &= E_{\textrm{BO}}(V)+ F_{\textrm{vib}}(V^{\bullet})\label{eq:Einfvib1}
+\Delta^{\bullet} V \frac{dF_{\textrm{vib}}}{dV}\Big|_{V^{\bullet},T}\\
F_{\textrm{E}\infty \textrm{Vib}2}(V,T) &= E_{\textrm{BO}}(V)+ F_{\textrm{vib}}(V^{\bullet})\label{eq:Einfvib2}
+\Delta^{\bullet} V \frac{ \partial F_{\textrm{vib}}}{ \partial V}\Big|_{V^{\bullet},T}
+\frac{1}{2} (\Delta^{\bullet} V)^2 \frac{ \partial ^2F_{\textrm{vib}}}{ \partial V^2}\Big|_{V^{\bullet},T},
\end{align}
where the first and second derivatives at each $T$ are computed using 2 and 3 points, respectively, as follows:
\begin{align}
\frac{dF_{\textrm{vib}}}{dV}\Big|_{V^{\bullet}}    &=\frac{ F_{\textrm{vib}}(V^{\bullet}+\Delta V)- F_{\textrm{vib}}(V^{\bullet}-\Delta V) }{2 \Delta V }\\
\frac{d^2F_{\textrm{vib}}}{dV^2}\Big|_{V^{\bullet}}&=\frac{ F_{\textrm{vib}}(V^{\bullet}+\Delta V)-2 F_{\textrm{vib}}(V^{\bullet})+ F_{\textrm{vib}}(V^{\bullet}-\Delta V) }{ \Delta V^2 }
\end{align}

 Likewise, to establish the equation for $F_{\textrm{E}{\infty} \textrm{Vib}4}(V,T)$ , 
 defined as :
\begin{align}
F_{\textrm{E}{\infty} \textrm{Vib}4}(V,T) =& E_{\textrm{BO}}(V)+ F_{\textrm{vib}}(V^{\bullet})\label{eq:Einfvib4}
+\Delta^{\bullet} V \frac{\partial F_{\textrm{vib}}}{\partial V}\Big|_{V^{\bullet},T}
+\frac{1}{2} (\Delta^{\bullet} V)^2 \frac{\partial ^2F_{\textrm{vib}}}{\partial V^2}\Big|_{V^{\bullet},T}\nonumber\\
+&\frac{1}{6} (\Delta^{\bullet} V)^3\frac{\partial ^3F_{\textrm{vib}}}{\partial V^3}\Big|_{V^{\bullet},T}
+\frac{1}{24} (\Delta^{\bullet} V)^4 \frac{\partial ^4F_{\textrm{vib}}}{\partial V^4}\Big|_{V^{\bullet},T}.
\end{align}
 It is necessary to define the derivative equations. Since a fourth-order derivative requires five data points, we utilize 5-point formulas to compute all derivatives.
\begin{align}
\frac{dF_{\textrm{vib}}}{dV  }  \Big|_{V^{\bullet}}&=\frac{-F_{\textrm{vib}}(V^{\bullet}+2\Delta V)+ 8 F_{\textrm{vib}}(V^{\bullet}+\Delta V)- 8  F_{\textrm{vib}}(V^{\bullet}-\Delta V)+  F_{\textrm{vib}}(V^{\bullet}-2 \Delta V) }{12\Delta V}\\ 
\frac{d^2F_{\textrm{vib}}}{dV^2}\Big|_{V^{\bullet}}&=\frac{-F_{\textrm{vib}}(V^{\bullet}+2\Delta V)+16 F_{\textrm{vib}}(V^{\bullet}+\Delta V)-30  F_{\textrm{vib}}(V^{\bullet})+16  F_{\textrm{vib}}(V^{\bullet}-\Delta V) -  F_{\textrm{vib}}(V^{\bullet}-2 \Delta V)}{12\Delta V^2}\\
\frac{d^3F_{\textrm{vib}}}{dV^3}\Big|_{V^{\bullet}}&=\frac{ F_{\textrm{vib}}(V^{\bullet}+2\Delta V)- 2 F_{\textrm{vib}}(V^{\bullet}+\Delta V)+ 2  F_{\textrm{vib}}(V^{\bullet}-\Delta V)-  F_{\textrm{vib}}(V^{\bullet}-2 \Delta V) }{2 \Delta V^3}\\
\frac{d^4F_{\textrm{vib}}}{dV^4}\Big|_{V^{\bullet}}&=\frac{ F_{\textrm{vib}}(V^{\bullet}+2\Delta V)- 4 F_{\textrm{vib}}(V^{\bullet}+\Delta V)+ 6  F_{\textrm{vib}}(V^{\bullet})- 4  F_{\textrm{vib}}(V^{\bullet}-\Delta V) +  F_{\textrm{vib}}(V^{\bullet}-2 \Delta V)}{\Delta V^4}
\end{align}

	\begin{figure*}[!htb]
		\includegraphics[width=0.55\textwidth]{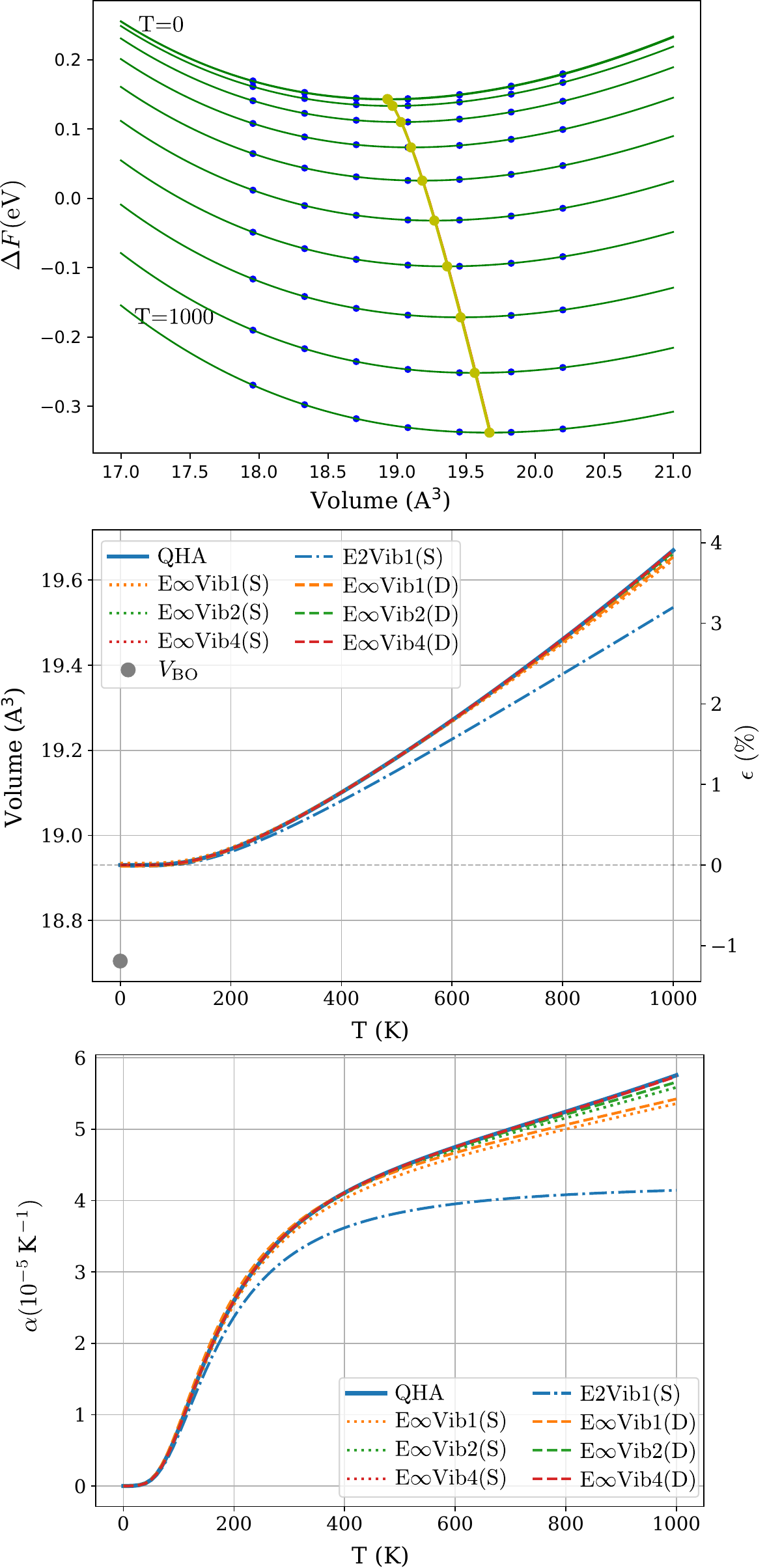}

  \caption{
 Detailed representation of free energy, volume, and 
 thermal expansion coefficient of MgO, for a temperature range from 0 K to 1000 K. In the top panel, free energies in QHA are charted. The minima of free energies are denoted by yellow points and lines. 
 The middle panel illustrates volume-temperature ($V$ vs. $T$) relationships for various approximations. Dashed lines represent results from the displaced selection of $V$, while dotted lines 
 depict results from symmetric selections. Solid lines denote QHA, and dot-dashed lines represent the E2Vib1 method. Right ticks indicate the percentage of volume change relative to the QHA volume at $T=0$, $\epsilon=(V-V_{\textrm{QHA}}(T=0))/V_{\textrm{QHA}}(T=0)$. The gray dot highlights the optimum value from BO DFT minimization, $V_{\textrm{BO}}$. In the bottom panel, thermal expansion coefficients are presented using different approximations, with the same line definitions as in the middle panel.}
		\label{fig:MgO}
	\end{figure*}
\begin{figure*}[!htb]
	\includegraphics[width=0.55\textwidth]{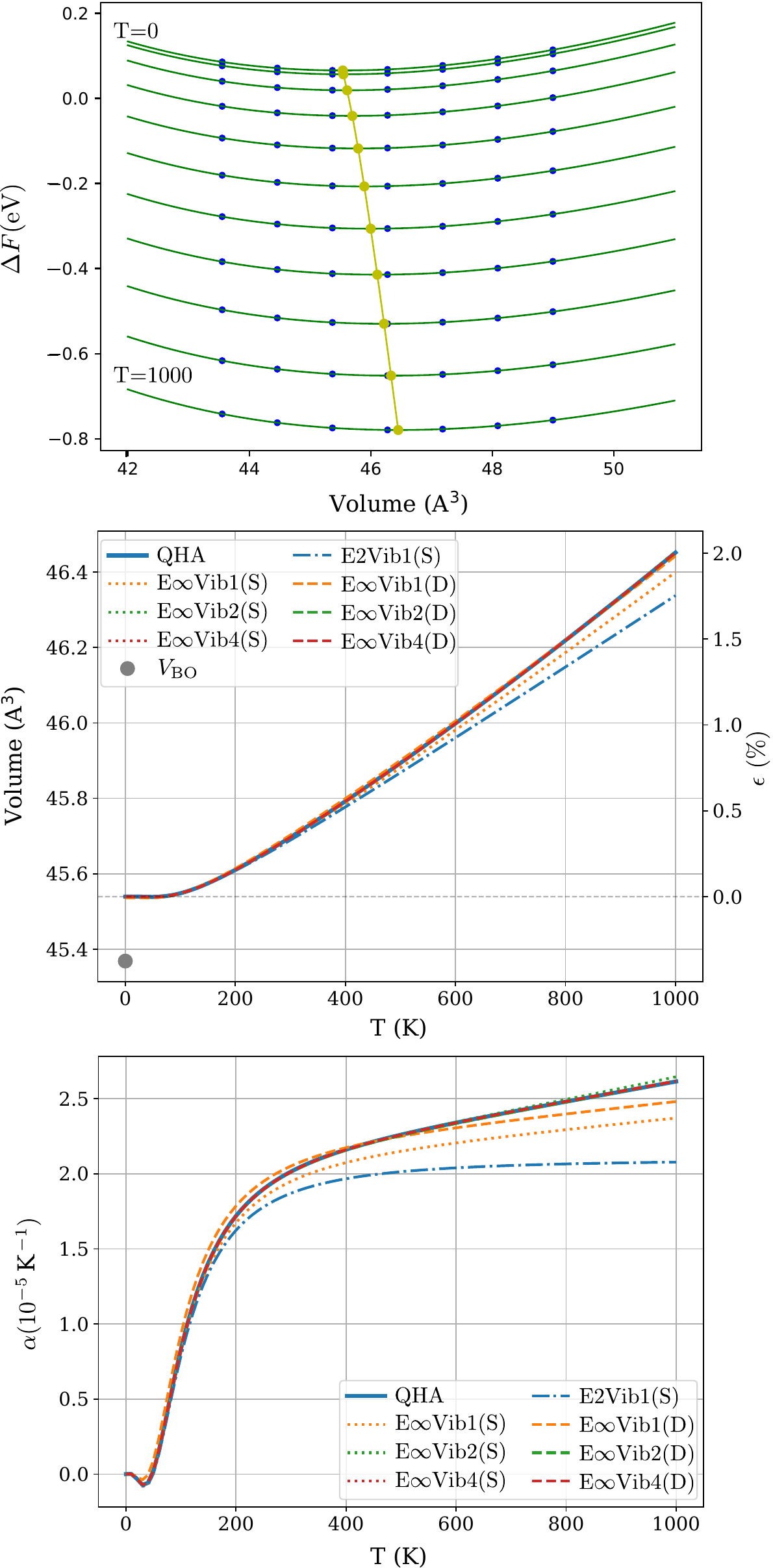}
	\caption{
		Comprehensive analysis of free Eenergies, volume-temperature 
  relationships, and thermal expansion coefficients of GaAs. Same conventions as Fig.~\ref{fig:MgO}.
	}
	\label{fig:GaAs}
\end{figure*}
\begin{figure*}[!htb]
	\includegraphics[width=0.55\textwidth]{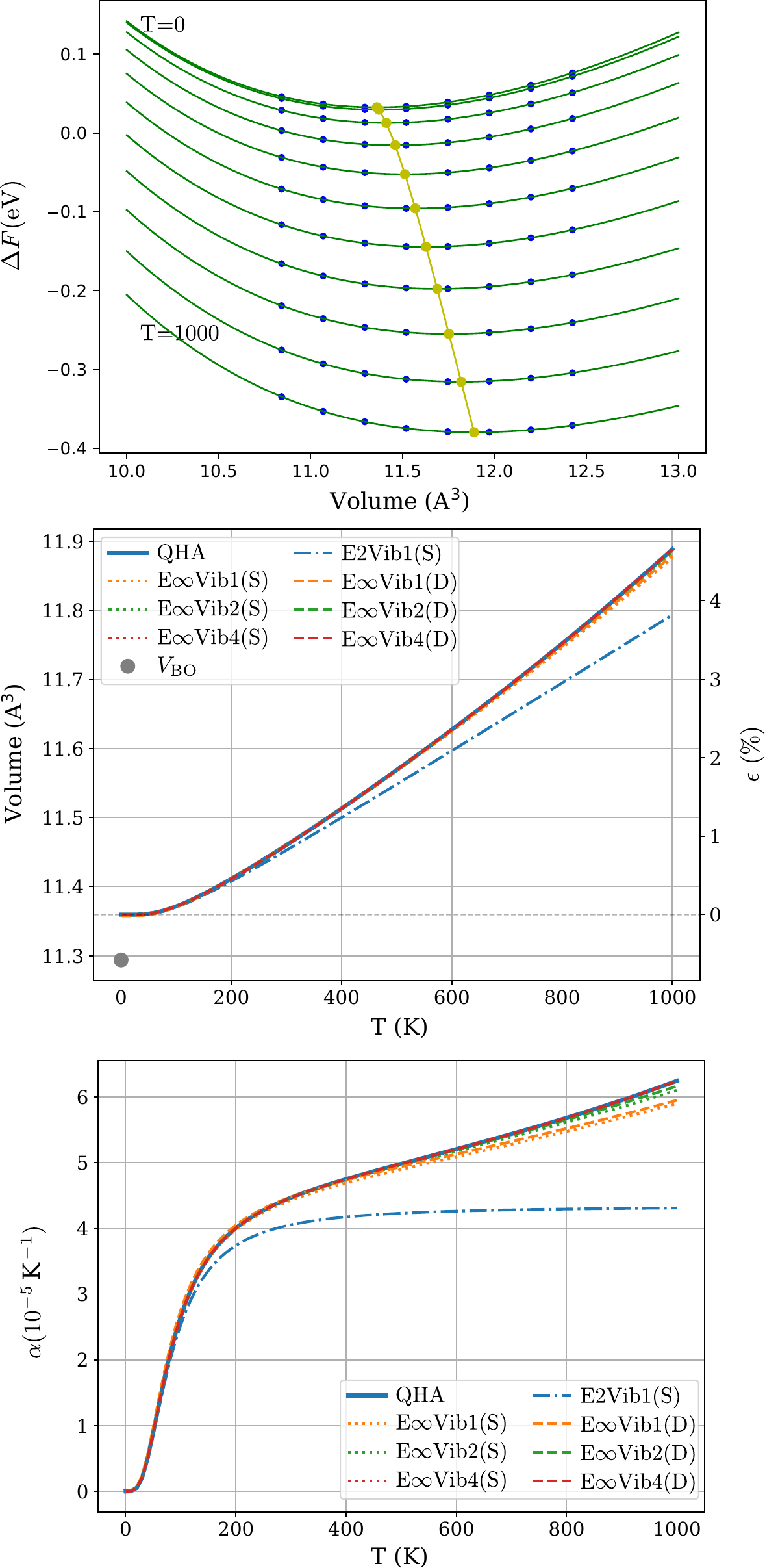}
	\caption{
		Comprehensive analysis of free energies, volume-temperature relationships, and thermal expansion coefficients of Cu. Same  conventions as Fig.~\ref{fig:MgO}.
	}
	\label{fig:Cu}
\end{figure*}
\begin{figure*}[!htb]
	\includegraphics[width=0.55\textwidth]{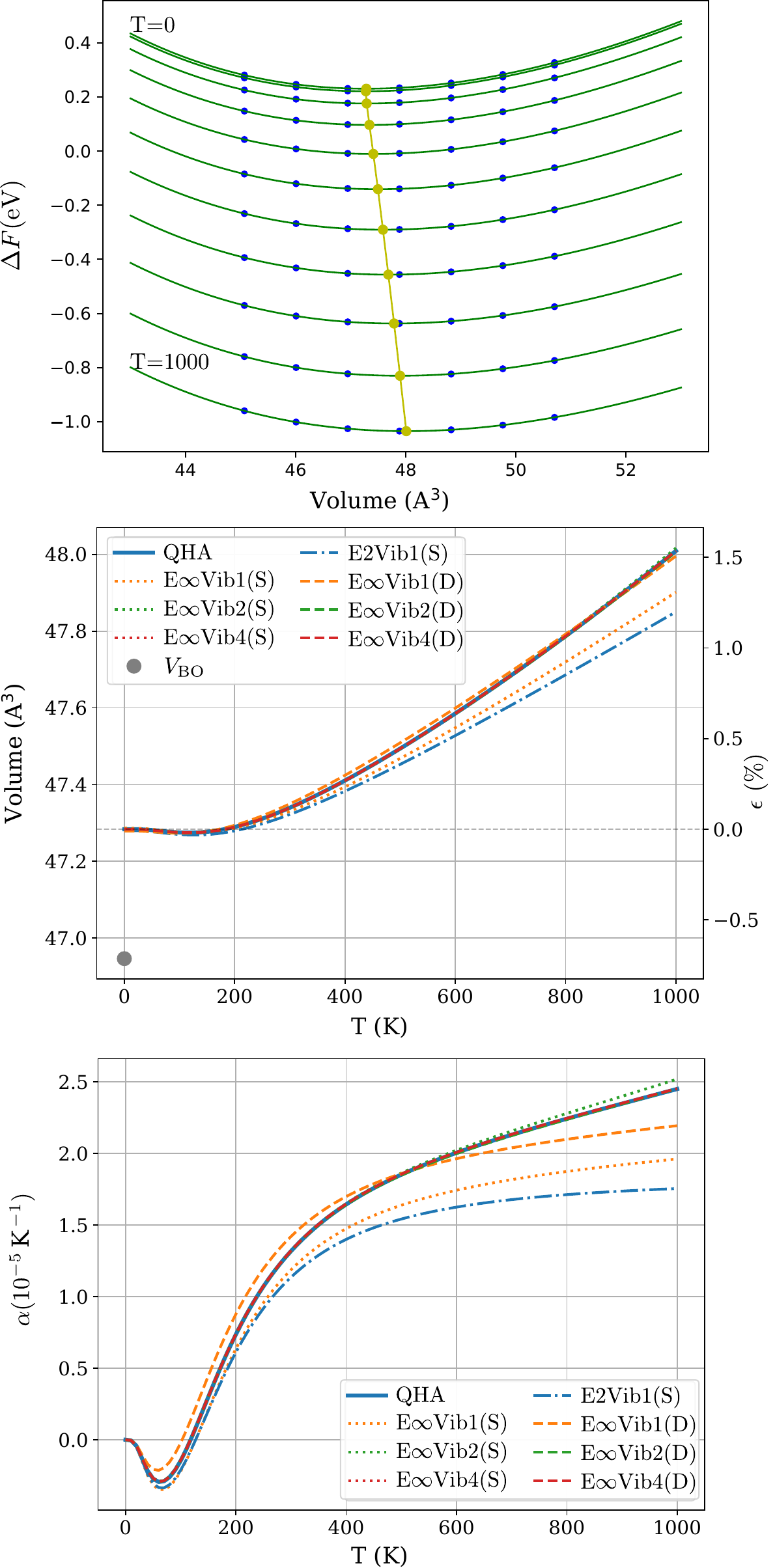}
	\caption{
		Comprehensive analysis of free energies, volume-temperature relationships, and thermal expansion coefficients of ZnO. Same  conventions as Fig.~\ref{fig:MgO}.
	}
	\label{fig:ZnO}
\end{figure*}
\begin{figure*}[!htb]
	\includegraphics[width=0.55\textwidth]{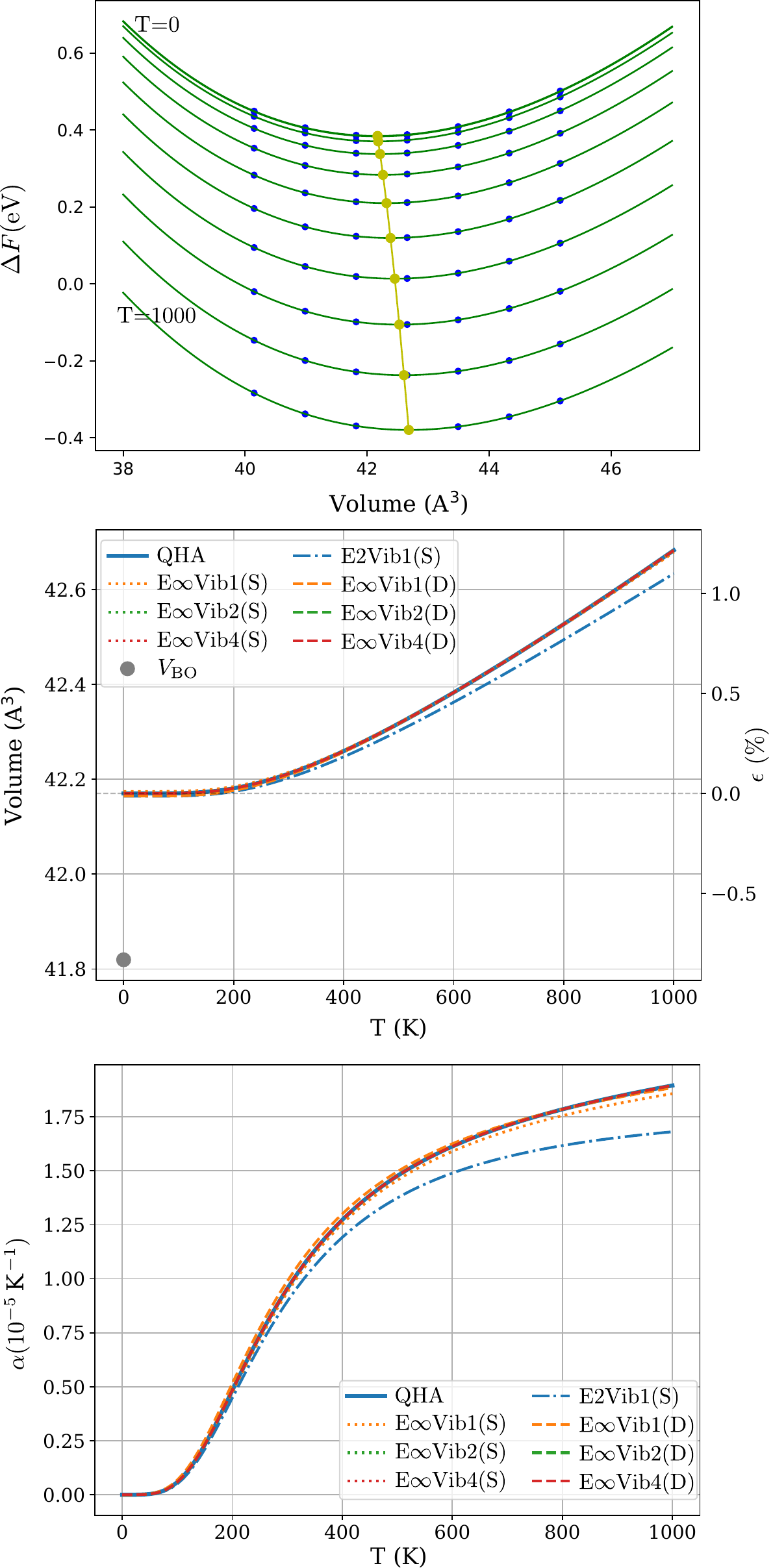}
	\caption{
		Comprehensive analysis of free energies, volume-temperature relationships, and thermal expansion coefficients of AlN. Same line conventions as Fig.~\ref{fig:MgO}.
	}
	\label{fig:AlN}
\end{figure*}
\begin{figure*}[!htb]
	\includegraphics[width=0.55\textwidth]{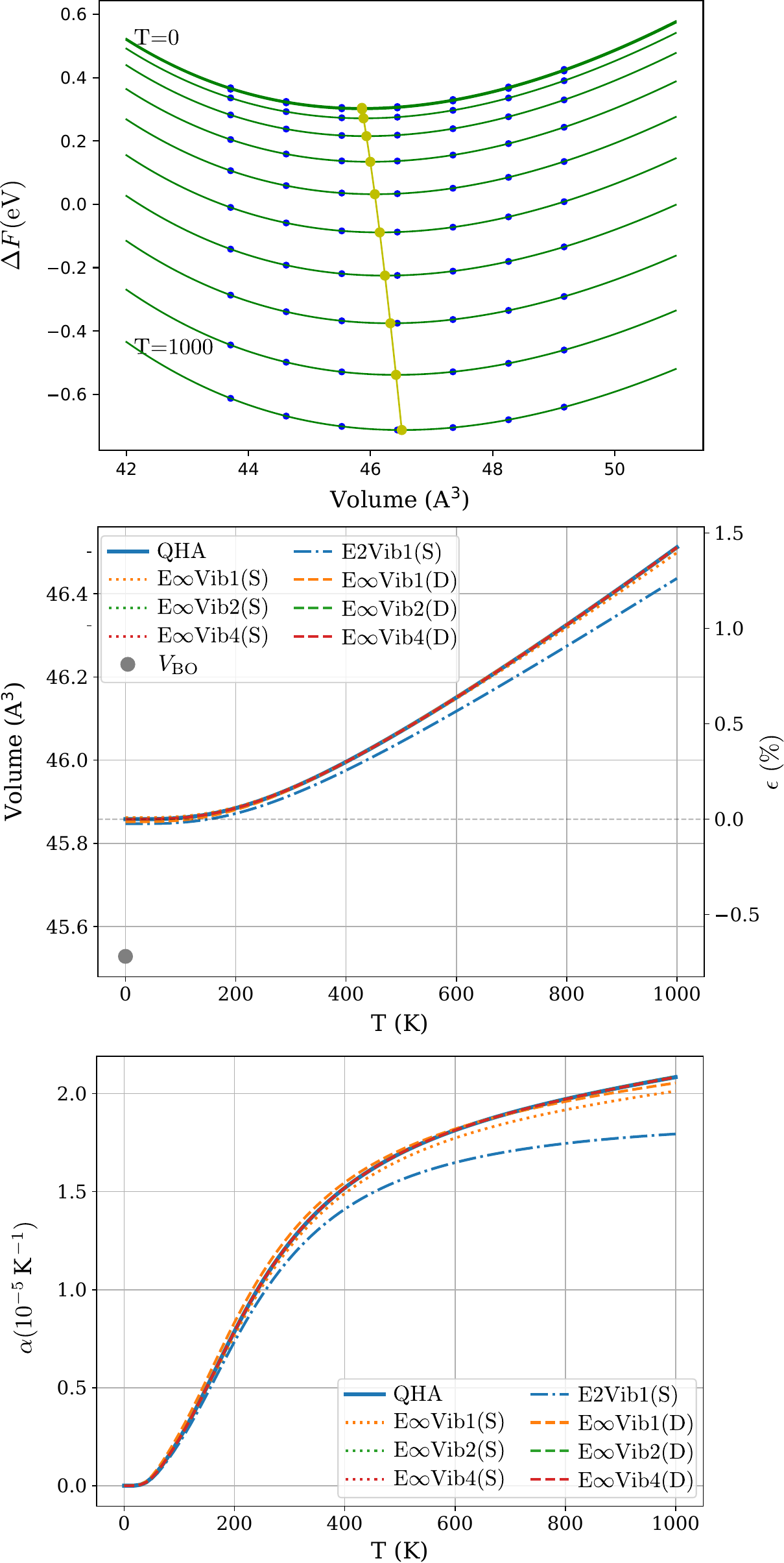}
	\caption{
		Comprehensive analysis of free energies, volume-temperature relationships, and thermal expansion coefficients of GaN. Same line conventions as Fig.~\ref{fig:MgO}.
	}
	\label{fig:GaN}
\end{figure*}
\begin{figure*}[!htb]
	\includegraphics[width=0.55\textwidth]{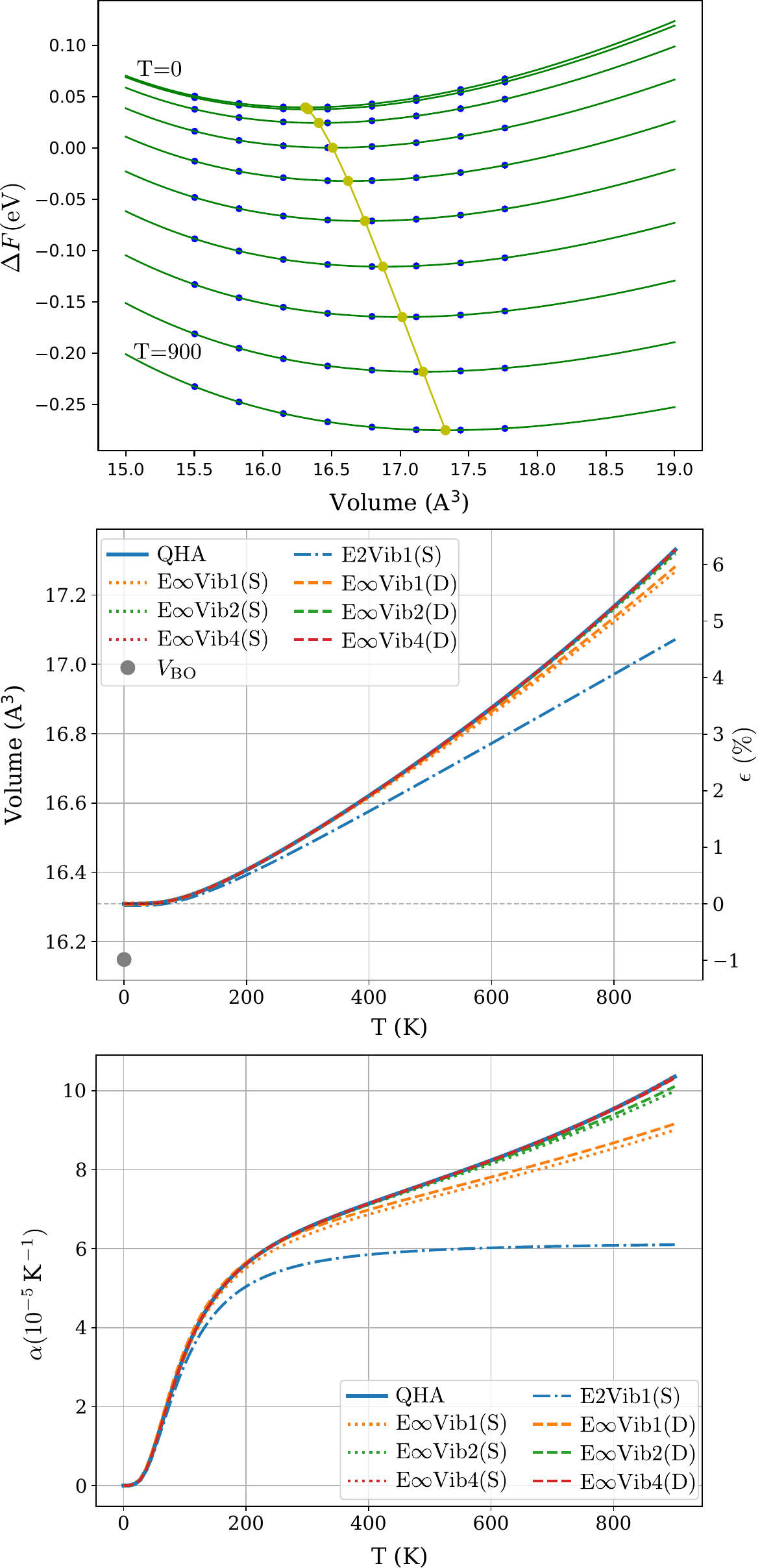}
	\caption{
		Comprehensive analysis of free energies, volume-temperature relationships, and thermal expansion coefficients of Al. Same line conventions as Fig.~\ref{fig:MgO}.
	}
	\label{fig:Al}
\end{figure*}
\begin{figure*}[!htb]
	\includegraphics[width=0.55\textwidth]{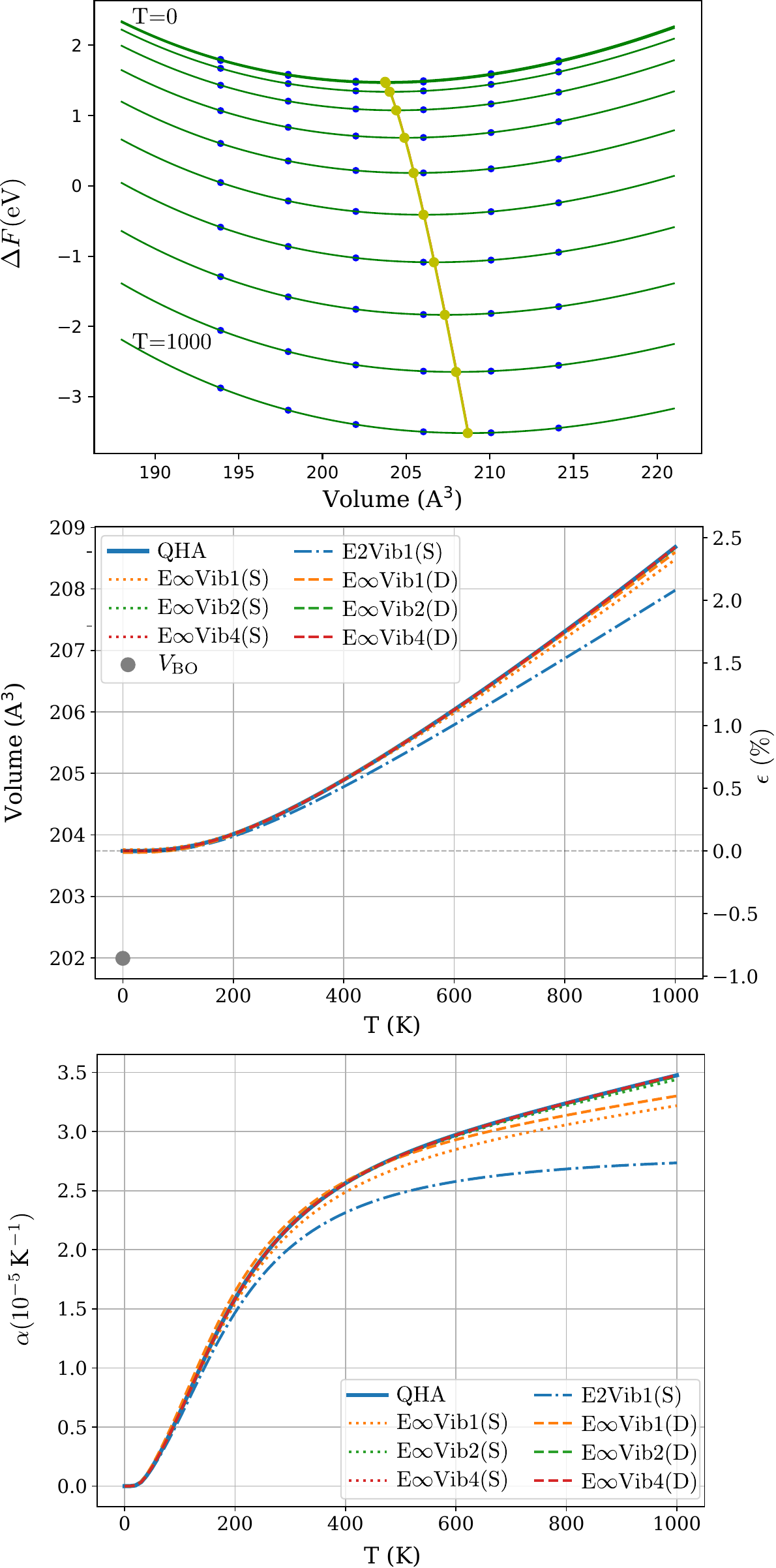}
	\caption{
		Comprehensive analysis of free energies, volume-temperature relationships, and thermal expansion coefficients of YAlO3. Same line conventions as Fig.~\ref{fig:MgO}.
	}
	\label{fig:YAlO3}
\end{figure*}
\begin{figure*}[!htb]
	\includegraphics[width=0.55\textwidth]{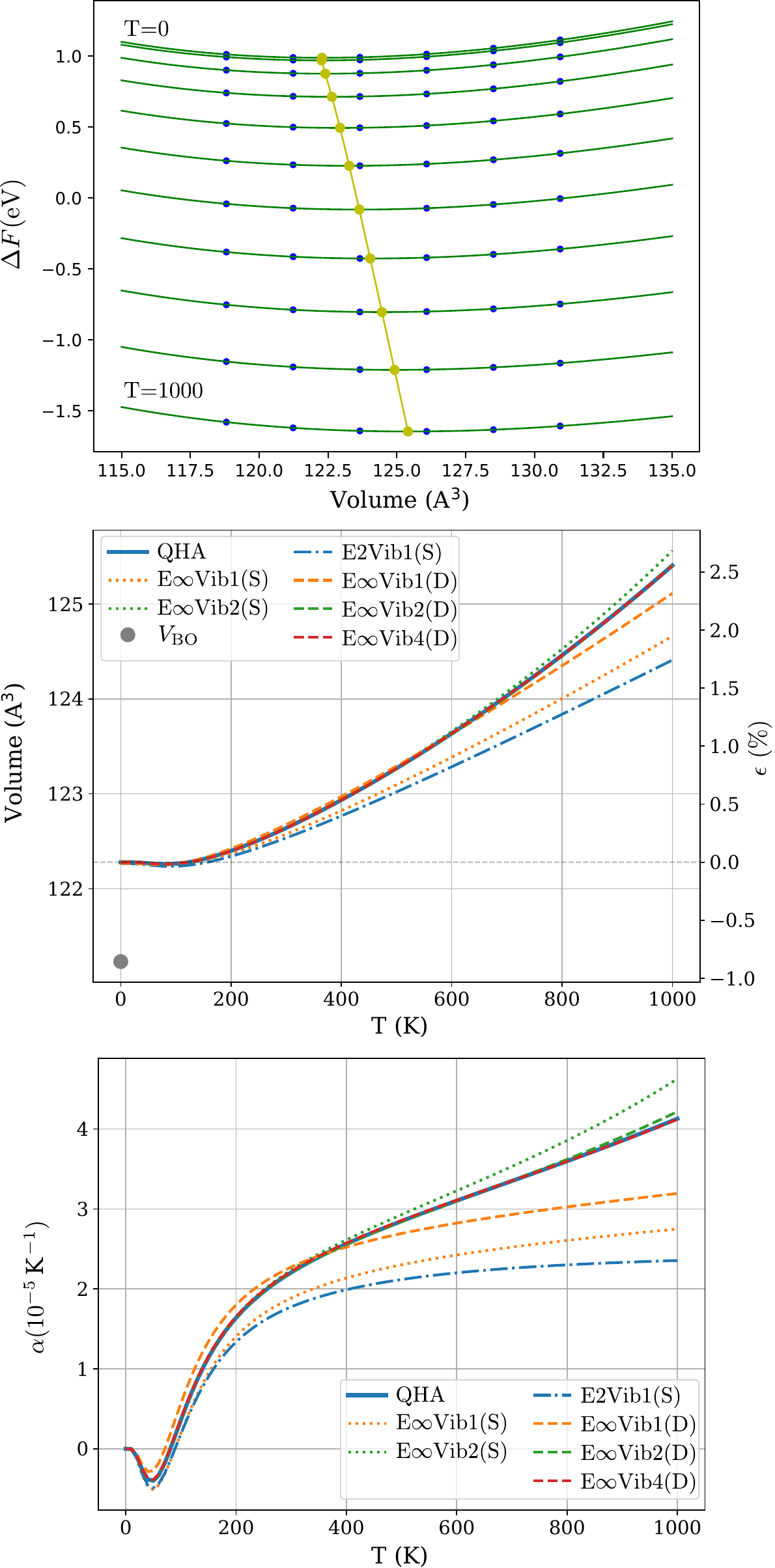}
	\caption{
		Comprehensive analysis of free energies, volume-temperature relationships, and thermal expansion coefficients of CaCO3. Same line conventions as Fig.~\ref{fig:MgO}.
	}
	\label{fig:CaCO3}
\end{figure*}
\newcommand{\vs}{\vspace{0.3em}}
\newcommand{\hs}{\hspace{0.5cm}}

\begin{table*}
    \caption{Lattice constants and reduced atomic coordinates for \ce{YAlO3}}
     \begin{tabular}{lccc}
~\\ 
     \hline
Lattice(\AA) :& 5.32162876  & \hs 0.00000000   & \hs   0.00000000  \vs \\
              & 0.00000000  & \hs 7.35809972   & \hs   0.00000000  \vs \\
              & 0.00000000  & \hs 0.00000000   & \hs   5.15852346  \vs \\
\hline
atom &x &y &z\\
\hline 
 Y  & 0.94488265 & \hs  0.75000000 & \hs 0.01315684 \vs \\
 Y  & 0.55511734 & \hs  0.25000000 & \hs 0.51315684 \vs \\
 Y  & 0.44488265 & \hs  0.75000000 & \hs 0.48684315 \vs \\
 Y  & 0.05511734 & \hs  0.25000000 & \hs 0.98684315 \vs \\
 Al & 0.49999999 & \hs  0.00000000 & \hs 0.00000000 \vs \\
 Al & 0.00000000 & \hs  0.00000000 & \hs 0.49999999 \vs \\
 Al & 0.00000000 & \hs  0.50000000 & \hs 0.49999999 \vs \\
 Al & 0.49999999 & \hs  0.50000000 & \hs 0.00000000 \vs \\
 O  & 0.29447495 & \hs  0.45428151 & \hs 0.70503708 \vs \\
 O  & 0.20552504 & \hs  0.54571848 & \hs 0.20503708 \vs \\
 O  & 0.79447495 & \hs  0.04571848 & \hs 0.79496291 \vs \\
 O  & 0.70552504 & \hs  0.95428151 & \hs 0.29496291 \vs \\
 O  & 0.70552504 & \hs  0.54571848 & \hs 0.29496291 \vs \\
 O  & 0.79447495 & \hs  0.45428151 & \hs 0.79496291 \vs \\
 O  & 0.20552504 & \hs  0.95428151 & \hs 0.20503708 \vs \\
 O  & 0.29447495 & \hs  0.04571848 & \hs 0.70503708 \vs \\
 O  & 0.47692059 & \hs  0.25000000 & \hs 0.08673512 \vs \\
 O  & 0.02307940 & \hs  0.75000000 & \hs 0.58673512 \vs \\
 O  & 0.97692059 & \hs  0.25000000 & \hs 0.41326487 \vs \\
 O  & 0.52307940 & \hs  0.75000000 & \hs 0.91326487 \vs \\
  \hline
    \end{tabular}

\end{table*}

\begin{table*}
    \caption{Lattice constants and reduced atomic coordinates for \ce{ZrO2}}
     \begin{tabular}{lccc}
~\\
     \hline
Lattice(\AA) :&  5.12569855 &\hs  0.00000000 & \hs 0.00000000  \vs \\
              &  0.00000000 &\hs  5.20922163 & \hs 0.00000000  \vs \\
              & -0.88120749 &\hs  0.00000000 & \hs 5.21920366  \vs \\
\hline
atom &x &y &z\\
\hline
  Zr & 0.27714492 & \hs 0.04311005 & \hs 0.20952312 \vs \\
  Zr & 0.27714492 & \hs 0.45688994 & \hs 0.70952312 \vs \\
  Zr & 0.72285507 & \hs 0.54311005 & \hs 0.29047687 \vs \\
  Zr & 0.72285507 & \hs 0.95688994 & \hs 0.79047687 \vs \\
  O  & 0.92982002 & \hs 0.83667720 & \hs 0.15868623 \vs \\
  O  & 0.92982002 & \hs 0.66332279 & \hs 0.65868623 \vs \\
  O  & 0.44876228 & \hs 0.74227000 & \hs 0.97962873 \vs \\
  O  & 0.55123771 & \hs 0.24227000 & \hs 0.52037126 \vs \\
  O  & 0.44876228 & \hs 0.75772999 & \hs 0.47962873 \vs \\
  O  & 0.55123771 & \hs 0.25772999 & \hs 0.02037126 \vs \\
  O  & 0.07017997 & \hs 0.33667720 & \hs 0.34131376 \vs \\
  O  & 0.07017997 & \hs 0.16332279 & \hs 0.84131376 \vs \\
  \hline
    \end{tabular}

\end{table*}

	

\end{document}